# Geometric Direction Finding on Dynamic Manifolds: Unambiguous DOA Estimation for Spatially Undersampled UWB Arrays

*Kailun Tian, Kaili Jiang, Dechang Wang, Hancong Feng, Yuxin Zhao, Ying Xiong, and Bin Tang*

***Abstract*—Traditional Direction of Arrival (DOA) estimation methods struggle to simultaneously address three physical constraints in Ultra-Wideband (UWB) electromagnetic sensing: spatial undersampling, asynchronous array phase, and beam squint. Existing solutions treat these issues in isolation, leading to limited performance in complex scenarios. This paper proposes a novel dynamic manifold perspective, which models UWB signal observations as a continuous manifold curve in a high-dimensional space driven by temporal evolution and array topology. We theoretically demonstrate that the DOA can be uniquely determined solely by the geometric shape of the manifold, rather than the absolute arrival phase. Based on this perspective, we construct a geometric parameter system comprising extrinsic and intrinsic parameters, along with a corresponding DOA estimation framework. Extrinsic vector parameters serve as a dynamic extension of traditional array processing, effectively expanding the degrees of freedom to suppress grating lobes. Intrinsic scalar invariants provide a new geometric perspective independent of traditional phase models, offering intrinsic robustness against array channel phase errors. Simulation results show that the derived analytical expressions for geometric parameters are highly consistent with numerical truths. The proposed framework not only completely eliminates spatial ambiguity in sparse arrays but also achieves high-precision direction finding under conditions with calibration-free phase errors.**

## I. INTRODUCTION

DRIVEN by the evolution of sixth-generation (6G) mobile communications [1] and Integrated Sensing and Communication (ISAC) technologies [2], Ultra-Wideband (UWB) [3] signal's Direction of Arrival (DOA) estimation increasingly relies on sparse arrays to overcome physical constraints [4]. However, coupling wideband signals with sparse topologies introduces a fundamental Space-Time-Phase ambiguity that disables traditional static processing. Specifically, the violation of the spatial sampling theorem generates dense grating lobes [5], while large bandwidths demanded by 6G induce severe beam squint [6]. Compounded by synchronization errors such as distributed nodes [7]. These multi-dimensional challenges render conventional narrowband algorithms based on static manifold assumptions ineffective for practical engineering.

Existing solutions typically treat the aforementioned difficulties as independent problems, addressing them separately [8].

Addressing the spatial ambiguity problem, sparse geometry designs (such as coprime [9] and nested arrays [10]) utilize the concept of difference co-arrays [11] to construct virtual uniform arrays. Although these methods successfully increase the degrees of freedom [12], they are typically based on narrowband signal models and cannot address wideband issues directly.

Regarding wideband DOA estimation, dominant frameworks include the Incoherent Signal-Subspace Method (ISM) [13] and the Coherent Signal-Subspace Method (CSM) [14]. ISM processes individual frequency bins independently [15] and often fails to resolve ambiguities when grating lobes overlap in the frequency domain. CSM employs focusing matrices to align signal subspaces from different frequencies to a reference frequency [16]. However, it relies on preliminary DOA estimates and is prone to focusing errors, especially when the sparse array itself exhibits inherent ambiguity. These methods tend to treat frequency variation as an error to be compensated for, rather than a feature to be utilized.

For phase errors, self-calibration algorithms (e.g., iterative optimization [17] or eigen structure-based methods [18]) have been developed to jointly estimate DOA and sensor phases [19]. However, these methods typically pose non-convex optimization problems [20], requiring initial guesses to avoid local minima. Furthermore, calibration performance degrades when the array manifold itself is ambiguous [21] due to sparsity.

In summary, current approaches operate within the framework of the static array manifold [22], lack a unified theoretical framework to exploit the dynamic information embedded in the continuous temporal evolution of non-stationary signals.

In this paper, we propose a novel dynamic manifold

This work was supported in part by the National Natural Science Foundation of China under Grant 62301119. *Corresponding author: Kailun Tian*.

Kailun Tian is with the University of Electronic Science and Technology of China, Chengdu, Sichuan, 611731, China (e-mail: kailun_tian@163.com).

Kaili Jiang is with the University of Electronic Science and Technology of China, Chengdu, Sichuan, 611731, China (jiangkelly@uestc.edu.cn).

Dechang Wang is with the University of Electronic Science and Technology of China, Chengdu, Sichuan, 611731, China (c13844033835@163.com).

Hancong Feng is with the University of Electronic Science and Technology of China, Chengdu, Sichuan, 611731, China (2927282941@qq.com).

Yuxin Zhao is with the University of Electronic Science and Technology of China, Chengdu, Sichuan, 611731, China (1051172535@qq.com).

Ying Xiong is with the University of Electronic Science and Technology of China, Chengdu, Sichuan, 611731, China (yxiong@uestc.edu.cn).

Bin Tang is with the University of Electronic Science and Technology of China, Chengdu, Sichuan, 611731, China (bint@uestc.edu.cn).

perspective, shifting the paradigm of DOA estimation from static phase matching to dynamic geometric evolution analysis. We model the array observation as a continuous trajectory evolving in a high-dimensional space. By introducing the dynamic generator algebra, we reveal a deep coupling between the signal's time-domain non-stationarity and the array geometry; this coupling drives the manifold to bend and twist.

Based on the differential geometry theory, we construct the Frenet-Serret moving frame [23] of the manifold and derive analytical expressions for the manifold's tangent vector, principal normal vector, binormal vector, curvature, and torsion. We prove that all these geometric features encode the DOA information. This perspective allows us to distinguish the true DOA based on the trajectory's geometric shape and evolution direction, rather than relying solely on ambiguous static phases.

The main contributions of this paper are summarized as follows. We provide the theoretical foundation for ambiguity-free DOA estimation using signal dynamics. We prove that the temporal dynamics of non-stationary signals can expand the observable signal subspace, thereby algebraically breaking the rank-1 ambiguity constraint of traditional narrowband models. Based on the dynamic manifold, ambiguity-free DOA estimation for wideband signals can be achieved under arbitrary array configurations. We classify the geometric parameters of the manifold into extrinsic parameters and intrinsic invariants. We prove that intrinsic parameters are invariant to rigid body transformations, thus providing the system with inherent robustness against uncalibrated phase errors. Based on the characteristics of geometric parameters, we offer two estimation frameworks. The framework I is based on extrinsic parameters, which utilizes complete trajectory information to achieve ambiguity resolution in arbitrary array systems. This serves as a dynamic extension of conventional array signal processing techniques. Framework II is based on intrinsic invariants, which is a geometric feature matching method that utilizes invariants to achieve robust, calibration-free estimation.

The remainder of this paper is organized as follows. Chapter II outlines the problem. Chapter III develops the differential geometric framework of the dynamic manifold. Chapter IV presents the theoretical analysis of observability and proposes the two general estimation frameworks. Chapter V provides comprehensive simulation results to validate the theory. Finally, Chapter VI concludes the paper.

## II. Problem Formulation

### A. Signal Model

Consider a sparse array consisting of $M$ omnidirectional elements. Let the $m$-th element position vectors be $\{\mathbf{p}_m\} \subset \mathbb{R}^3$. Assume the far-field signal $s(t) = A(t)e^{j\Phi(t)}$ impinges on the array from a direction $\mathbf{d}(\theta) \in \mathbb{R}^3$, where $\theta$ denotes the DOA. Dynamic means the instantaneous angular frequency $\omega(t)$ of this signal model is not constant.

$$\omega(t) = \frac{d\Phi(t)}{dt} \tag{1}$$

where $\Phi(t)$ is the phase of $s(t)$. Assume the envelope of $s(t)$ slowly varying $A(t-\tau) \approx A(t)$. The received signal at the $m$-th element can be expressed as a delayed version of $s(t)$

$$x_m(t) = s(t - \tau_m(\theta)) + n_m(t) \tag{2}$$

where $\tau_m(\theta) = \mathbf{p}_m^T \mathbf{d}(\theta)/c$ represents the propagation delay, $n_m(t)$ is white complex Gaussian noise, and $c$ is the speed of light. Substituting the above equation into the signal model

$$x_m(t) = s(t)a_m(\theta,t) + n_m(t)\,. \tag{3}$$

The vector $\mathbf{a}(\theta,t) = [a_1(\theta,t),...,a_M(\theta,t)]^T$, where $a_m(\theta,t) = e^{j(\Phi(t-\tau_m(\theta))-\Phi(t))}$, is referred to as the steering vector. Thus, the observation vector $\mathbf{x}(t) \in \mathbb{C}^M$ is

$$\mathbf{x}(t) = \begin{bmatrix} x_1(t) \\ \vdots \\ x_M(t) \end{bmatrix} = s(t)\mathbf{a}(\theta,t) + \mathbf{n}(t)\,. \tag{4}$$

For the sake of simplicity, we will introduce noise only where necessary in the subsequent discussion.

### B. Three Fundamental Challenges

There are three fundamental challenges that cause conventional DOA estimation methods to become ineffective.

First, in typical array signal processing, the signal frequency is typically assumed to be constant $\omega(t) = 2\pi f_c$. In this case, $a_m(\theta,t)$ degenerates into the standard form $a_m(\theta)$.

$$a_m(\theta) = e^{j\left(\int_0^{t-\tau_m(\theta)} \omega(\xi)d\xi - \int_0^t \omega(\xi)d\xi\right)} = e^{-j2\pi f_c \tau_m(\theta)} \tag{5}$$

Ideally, this phase has a one-to-one correspondence with DOA. However, spatial undersampling leads to phase ambiguity. For high carrier frequency signals, the inter-element spacing of sparse arrays is usually much larger than half the minimum wavelength ( $d \gg \lambda_{\min}/2$ ). Consequently, the measured phase difference $\Delta\phi_{meas}$ between any two sensors $m$ and $n$ contains an integer cycle ambiguity

$$\Delta\phi_{meas}(t) = \left[-2\pi f_c(\tau_m(\theta) - \tau_n(\theta))\right]_{2\pi}\,. \tag{6}$$

Which can be rewritten as

$$-2\pi f_c(\tau_m(\theta) - \tau_n(\theta)) = \Delta\phi_{meas}(t) + 2k\pi, k \in \mathbb{Z}\,. \tag{7}$$

Since the integer $k$ is unknown, multiple solutions for $\theta$ exist. This phenomenon creates grating lobes and results in direction-finding ambiguity.

Second, frequency non-stationarity leads to beam squint. Using the mean value theorem, the phase difference term can be expressed as

$$\Phi(t - \tau_m(\theta)) - \Phi(t) = -\int_{t-\tau_m}^{t} \omega(\xi)d\xi \approx -\omega(t)\tau_m(\theta)\,. \tag{8}$$

Considering $\omega(t)$ is not constant, $a_m(\theta,t) \approx e^{-j\omega(t)\tau_m(\theta)}$, time $t$ is explicitly incorporated. This implies that for broadband signals, the steering vector is no longer a static geometric vector. Time-varying steering vector causes mismatches in conventional signal processing theory.

Finally, practical array systems often suffer from uncalibrated inter-element phase errors.

$$\mathbf{\Gamma} = \mathrm{diag}(e^{j\phi_1},\ldots,e^{j\phi_M}) \tag{9}$$

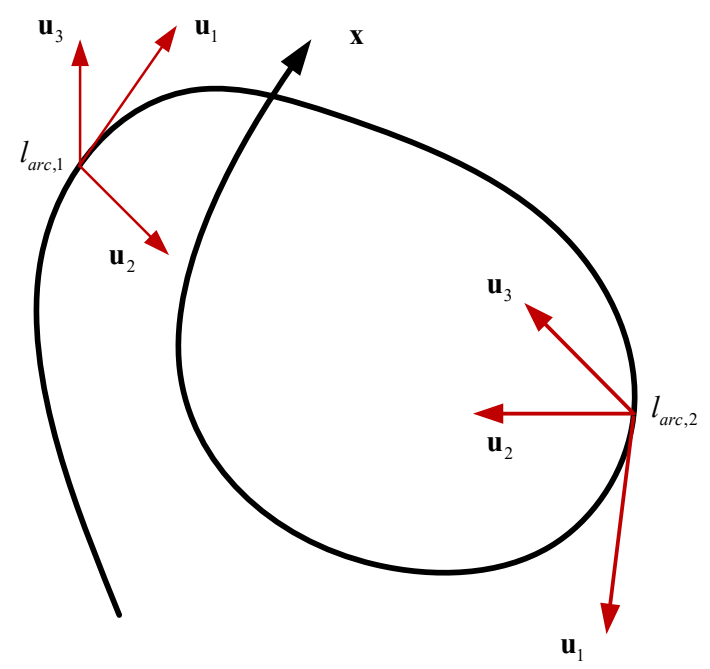

**Fig. 1.** Schematic diagram of the Frenet-Serret frame

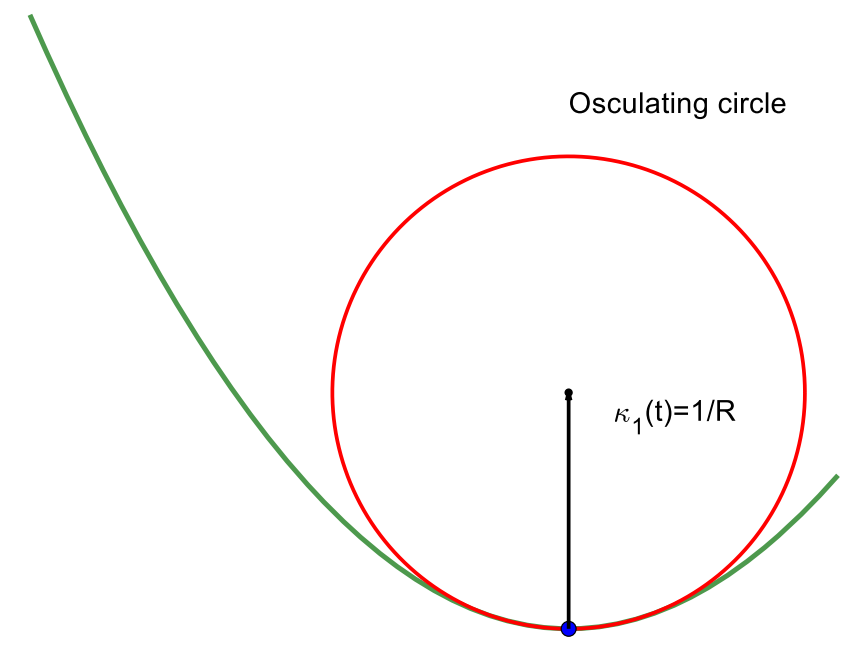

**Fig. 2. Schematic Diagram of Curvature**

These phase errors induce a coordinate rotation of the steering vector. This rotation causes a severe mismatch between the received data and the theoretical model, thereby rendering conventional DOA estimation algorithms ineffective.

In summary, the steering vector is a key parameter for current array processing techniques. However, spatial undersampling, signals with dynamic models or phase errors cause DOA algorithms based on the steering vector ineffective.

## III. Dynamic Manifold and Differential Geometry

### *A. Dynamic Manifold of Array Observations*

The analysis in Section 2 indicates that traditional array signal processing is no longer applicable. It is necessary to introduce time dependency to address the modeling mismatch.

For a fixed $\theta$, the set of time delays $\{\tau_m(\theta)\}$ consists of fixed constants. Therefore, $\mathbf{x}(t)$ is equivalent to observing the same scalar system $s(t)$ with multiple time delays. The evolution of $\mathbf{x}(t)$ over time $t$ traces a continuous trajectory in the high-dimensional complex space. This trajectory constitutes the dynamic manifold $\mathcal{M}_{obs}$, which is different from the array manifold $A_{array} = span\{a(\theta) \mid \theta \in (-\pi, \pi]\}$.

$$\mathcal{M}_{obs}(t) = \left\{ \mathbf{x}(t) \in \mathbb{C}^M \mid t \in T_{obs} \right\}. \tag{10}$$

Parameterized by the physical time $t$, $\mathcal{M}_{obs}$ is a one-dimensional smooth trajectory embedded in the complex Euclidean space $\mathbb{C}^M$. It can be seen that, time $t$ determining the position of the manifold, while $\theta$ degenerates into the structural parameter of the manifold. Clearly, the trajectory is a curve. For consistency, all subsequent references to the manifold refer to the observation dynamic manifold $\mathcal{M}_{obs}$.

The geometric analysis in this paper focuses on the continuous evolution intervals where the trajectory is differentiable ( $C^\infty$ ), such as Linear Frequency Modulation (LFM) and Sinusoidal Frequency Modulation (SFM) signals. The complex envelope $s(t)$ and its phase function $\Phi(t)$ are continuous and differentiable with respect to time $t$. The symbol transitions (singularities) are treated as boundaries of these smooth segments, such as frequency-shift or phase-shift keying signals, we model them as piecewise smooth curves. Since the array manifold response is also a smooth function, the smoothness of observation dynamic manifold $\mathcal{M}_{obs}$ is ensured.

Furthermore, we examine the topological constraints of the manifold. Assuming the source signal is amplitude-normalized and the modulus of the array manifold vector components is unity, the Euclidean norm of the observation vector satisfies

$$\|\mathbf{x}(t)\|^2 = \sum_{m=1}^{M} |x_m(t)|^2 = M . \tag{11}$$

This implies that, regardless of how the signal's dynamic characteristics change, the manifold $\mathcal{M}$ is always strictly confined to a Hypersphere of radius $\sqrt{M}$ in the complex Euclidean space $\mathbb{C}^M$.

To quantitatively describe the local geometric properties of the manifold curve, we need to establish a local coordinate system that moves along the curve. Given the isomorphism between the observation space $\mathbb{C}^M$ and the real Riemannian manifold $\mathbb{R}^{2M}$ , we adopt the concept of Wide-Sense Orthogonality [22]. Two complex vectors $\mathbf{u}, \mathbf{w} \in \mathbb{C}^M$ are defined as generalized orthogonal if and only if the real part of their Hermitian inner product is zero.

$$\mathrm{Re}\langle \mathbf{u}, \mathbf{w} \rangle = \mathrm{Re}(\mathbf{u}^H \mathbf{w}) = 0 \tag{12}$$

Based on this, we can construct a set of generalized orthogonal unit basis vectors $\{\mathbf{u}_1, \mathbf{u}_2, \ldots, \mathbf{u}_{2M}\}$ for the manifold. These vectors constitute the Frenet-Serret frame describing the dynamic evolution of the manifold, as shown in Fig. 1. To obtain features that depend only on the geometric shape of the manifold, we introduce the Arc Length $s$ (denoted as $l_{arc}$ to distinguish from the signal $s(t)$ ) as the natural parametrization of the curve.

For an example, according to the fundamental theorem of differential geometry, the rate of change of the tangent vector's direction $\mathbf{u}_1$ with respect to the arc length $l_{arc}$ defines the curvature $\kappa_1$. The curvature of a one-dimensional curve can be represented by the osculating circle, as shown in Fig. 2.

The subsequent sections will derive the specific mapping relationship between this geometric and the physical parameters.

*B. Physical Differential Vector and the Dynamic Generator*

This section introduces physical time $t$ as the dynamic parameter of the manifold, derives the differential form of the observation equation.

Define the physical velocity vector $\mathbf{v}(t) = d\mathbf{x}(t)/dt$ as the first derivative of the observation manifold $\mathbf{x}(t)$. For the $m$-th sensor, $v_m(t) = ds(t-\tau_m(\theta))/dt$. Under the slowly varying envelope assumption, we have

$$\dot{s}(t) = \frac{d}{dt}e^{j\Phi(t)} = j\dot{\Phi}(t)e^{j\Phi(t)} = j\omega(t)s(t) \tag{13}$$

Applying the chain rule shows that the physical velocity vector contains not only the time evolution information of the signal but also couples the array's geometry.

$$v_m(\theta,t) = j\omega(t-\tau_m(\theta))x_m(t) \tag{14}$$

The complete physical velocity vector is

$$\begin{aligned}\mathbf{v}(\theta,t) &= \mathrm{diag}(j\omega(t-\tau_1(\theta)),\ldots,j\omega(t-\tau_M(\theta)))\cdot\mathbf{x}(t)\\ &= \mathbf{\Omega}(\theta,t)\cdot\mathbf{x}(t)\end{aligned}. \tag{15}$$

Here, $\mathbf{\Omega}(\theta,t)$ is referred to as the dynamic generator, as it governs the evolution of the manifold induced by source dynamics and array geometry. This equation shows that $\mathbf{v}(t)$ satisfies a closed first-order linear differential system. The linear transformation relationship between $\mathbf{v}$ and $\mathbf{x}$ encodes the spatial parameter $\theta$.

We can perform a Taylor expansion on the instantaneous frequency $\omega(t-\tau_m)$ based on the smooth assumption.

$$\omega(t-\tau_m) = \sum_{n=0}^{\infty}\frac{(-1)^n}{n!}\omega^{(n)}(t)\cdot\tau_m^n \tag{16}$$

Consequently, the dynamic generator can be expanded

$$\mathbf{\Omega}(\theta,t) = \sum_{n=0}^{\infty}\mathbf{\Omega}_n = j\left[\omega(t)\mathbf{I} + \sum_{n=1}^{\infty}\frac{(-1)^n}{n!}\omega^{(n)}(t)\mathbf{T}^n(\theta)\right] \tag{17}$$

where $\mathbf{T}(\theta) = \mathrm{diag}(\tau_1(\theta),\tau_2(\theta),...,\tau_M(\theta))$ is the time-delay matrix. In (17), the zeroth-order term $j\omega(t)\mathbf{I}$ represents the common-mode dynamic evolution, dominated by the instantaneous frequency. In contrast, the higher-order terms represent the differential geometric perturbations induced by the coupling between the signal's dynamic and the array delay.

The second-order derivative is acceleration vector $\mathbf{a}_{cc}(t) = \ddot{\mathbf{x}}(t)$, which can be driven by

$$\mathbf{a}_{cc}(t) = \frac{d}{dt}(\mathbf{\Omega}\mathbf{x}) = (\dot{\mathbf{\Omega}} + \mathbf{\Omega}^2)\mathbf{x}. \tag{18}$$

Similarly, the third-order derivative is jerk vector $\mathbf{j}(t) = \dddot{\mathbf{x}}(t)$

$$\mathbf{j}(t) = (\ddot{\mathbf{\Omega}} + 2\dot{\mathbf{\Omega}}\mathbf{\Omega} + \mathbf{\Omega}\dot{\mathbf{\Omega}} + \mathbf{\Omega}^3)\mathbf{x}. \tag{19}$$

By analogy, all high-order physical derivatives $\mathbf{x}^{(k)}$ of the manifold can be recursively generated by $\mathbf{\Omega}$. The geometric properties of the manifold (such as bending and twisting) depend on the angles between these derivative vectors. To analyze these angles, we introduce the following core lemma.

**Lemma 1**. In an ideal lossless array, the dynamic operator $\mathbf{\Omega}(\theta,t)$ is an Anti-Hermitian Operator, satisfying $\mathbf{\Omega}^H = -\mathbf{\Omega}$.

Since the modulus of each element in the manifold is constant, its phase $\Phi_m(t)$ is a real function. Therefore, the diagonal elements of $\mathbf{\Omega}$, are purely imaginary, i.e., $\mathbf{\Omega}^H = -\mathbf{\Omega}$.

The essence of this property is that $\mathbf{\Omega}$ represents a rotation on the high-dimensional complex hypersphere. Just as skew-symmetric matrices generate rotations in real space, anti-Hermitian operators generate unitary evolution in complex space, ensuring the manifold remains constrained on the hypersphere. Based on Lemma 1, within the framework of real embedded manifold geometry, we can derive a series of orthogonality relationships between physical derivatives.

**Corollary 1**. If the operator $\mathbf{\Omega}$ is anti-Hermitian, then for odd term $\mathbf{\Omega}^{2k+1}\mathbf{x}$ and even term $\mathbf{\Omega}^{2m}\mathbf{x}$, the results of its operator powers satisfy $\langle\mathbf{\Omega}^{2k+1}\mathbf{x},\mathbf{\Omega}^{2m}\mathbf{x}\rangle_{\mathbb{R}} = 0$.

For an example, the physical velocity vector $\mathbf{v}$ is always perpendicular to the observation vector

$$\langle\mathbf{x},\mathbf{v}\rangle_{\mathbb{R}} = \mathrm{Re}(\mathbf{x}^H\mathbf{\Omega}\mathbf{x}) = 0. \tag{20}$$

Through this algebraic framework, we transform complex differential geometric operations into clear operator algebraic operations.

*C. Geometric Parameter Conversion and Frenet-Serret Frame*

This section builds the bridge between the geometric Frenet-Serret frame and the physical time domain.

First, we need to establish the metric conversion between the natural parameter $l_{arc}$ and time $t$. According to the definition of the arc length $dl_{arc} = \|d\mathbf{x}\|$, we can derive

$$\frac{dl_{arc}}{dt} = \frac{\|d\mathbf{x}\|}{dt} = \left\|\frac{d\mathbf{x}}{dt}\right\| = \|\mathbf{v}(t)\|. \tag{21}$$

This factor reveals an extremely important physical meaning: the rate of change of arc length of the manifold curve is directly determined by the magnitude of the physical velocity vector. Based on the precise analysis of the tangent vector, we can also derive the exact arc length formula for the dynamic manifold. The exact magnitude of the tangent vector is given by

$$\|\mathbf{v}(t)\|^2 = \sum_{m=1}^{M}|j\omega(t-\tau_m)x_m(t)|^2 = \sum_{m=1}^{M}\omega^2(t-\tau_m) \tag{22}$$

This formulation provides the exact $l_{arc}$ for arbitrary signals.

$$l_{arc}(t,\theta) = \int_0^t\sqrt{\sum_{m=1}^{M}\omega^2(\xi-\tau_m(\theta))}\,d\xi \tag{23}$$

Different $\theta$ result in different $\tau_m(\theta)$, which fine-tunes the stretching speed of the curve. According to the definition in differential geometry, the unit tangent vector $\mathbf{u}_1(l_{arc})$ is the derivative of the position vector with respect to the arc length.

$$\mathbf{u}_1(l_{arc}) = \frac{d\mathbf{x}}{dl_{arc}} \tag{24}$$

Using the chain rule, we have

$$\mathbf{u}_1(t) = \frac{d\mathbf{x}/dt}{dl_{arc}/dt} = \frac{\mathbf{v}(t)}{\|\mathbf{v}(t)\|} = \frac{\mathbf{\Omega}\mathbf{x}}{\|\mathbf{\Omega}\mathbf{x}\|}. \tag{25}$$

This equation reveals that the direction of $\mathbf{u}_1$ is entirely determined by the direction of $\mathbf{v}$. This implies that the evolution direction of the dynamic generator directly indicates the geometric extension direction of the manifold.

The principal normal vector $\mathbf{u}_2$ is defined as the normalized direction of the derivative of the tangent vector

$$\mathbf{u}_2(l_{arc}) = \frac{\mathbf{u}_1'(l_{arc})}{\|\mathbf{u}_1'(l_{arc})\|} \tag{26}$$

Using the chain rule

$$\frac{d\mathbf{u}_1}{dl_{arc}} = \frac{d\mathbf{u}_1/dt}{dl_{arc}/dt} = \frac{1}{\|\mathbf{v}\|}\frac{d}{dt}\left(\frac{\mathbf{v}}{\|\mathbf{v}\|}\right) = \frac{1}{\|\mathbf{v}\|}\frac{\dot{\mathbf{v}}\|\mathbf{v}\| - \mathbf{v}\dfrac{d\|\mathbf{v}\|}{dt}}{\|\mathbf{v}\|^2}. \tag{27}$$

Note that $\dot{\mathbf{v}}(t) = \ddot{\mathbf{x}}(t) = \mathbf{a}_{cc}(t)$ is the physical acceleration vector, and the physical velocity vector $\mathbf{v}$ can be decomposed into magnitude ($\|\mathbf{v}\|$) multiplied by unit direction ($\mathbf{u}_1$).

$$\mathbf{a}_{cc} = \frac{d}{dt}(\|\mathbf{v}\|\mathbf{u}_1) = \underbrace{\frac{d\|\mathbf{v}\|}{dt}\mathbf{u}_1}_{\mathbf{a}_{\tan}} + \underbrace{\|\mathbf{v}\|\frac{d\mathbf{u}_1}{dt}}_{\mathbf{a}_{\text{norm}}} \tag{28}$$

The second term of numerator in the $\dfrac{d\mathbf{u}_1}{dl_{arc}}$ is actually the component of acceleration in the tangential direction (tangential acceleration $\mathbf{a}_{\tan}$).

$$\mathbf{v}\frac{d\|\mathbf{v}\|}{dt} = \|\mathbf{v}\|\mathbf{u}_1\frac{d\|\mathbf{v}\|}{dt} = \|\mathbf{v}\|\mathbf{a}_{\tan} \tag{29}$$

Therefore, the numerator in $\dfrac{d\mathbf{u}_1}{dl_{arc}}$ as a whole represents the normal acceleration after subtracting the tangential component.

$$\frac{d\mathbf{u}_1}{dl_{arc}} = \frac{1}{\|\mathbf{v}\|}\frac{(\mathbf{a}_{\tan} + \mathbf{a}_{\text{norm}})\|\mathbf{v}\| - \|\mathbf{v}\|\mathbf{a}_{\tan}}{\|\mathbf{v}\|^2} = \frac{\mathbf{a}_{\text{norm}}}{\|\mathbf{v}\|^2} \tag{30}$$

Clearly, the direction of $\mathbf{u}_2$ is perpendicular to $\mathbf{u}_1$. We can extract this direction by projecting onto the orthogonal complement of the tangent space. Introducing the normal projection operator $\mathbf{P}_{\mathbf{v}}^{\perp} = \mathbf{I} - \mathbf{u}_1\mathbf{u}_1^H$. Using $c_1$ as the coefficient.

$$c_1 = \langle \mathbf{u}_1, \mathbf{a}_{cc} \rangle_{\mathbb{R}} = \text{Re}\left(\frac{(\mathbf{\Omega x})^H(\dot{\mathbf{\Omega}} + \mathbf{\Omega}^2)\mathbf{x}}{\|\mathbf{\Omega x}\|}\right) \tag{31}$$

According to corollary 1

$$\text{Re}\left((\mathbf{\Omega x})^H(\mathbf{\Omega}^2\mathbf{x})\right) = 0 \tag{32}$$

$$c_1 = \text{Re}\left(\frac{-\mathbf{x}^H\mathbf{\Omega}\dot{\mathbf{\Omega}}\mathbf{x}}{\|\mathbf{\Omega x}\|}\right) \tag{33}$$

We can derive the analytical expression for $\mathbf{u}_2(t)$.

$$\mathbf{P}_{\mathbf{v}}^{\perp}\mathbf{a}_{cc} = \mathbf{a}_{cc} - c_1\mathbf{u}_1 = (\dot{\mathbf{\Omega}} + \mathbf{\Omega}^2)\mathbf{x} - c_1\frac{\mathbf{\Omega x}}{\|\mathbf{\Omega x}\|} \tag{34}$$

$$\mathbf{u}_2(t) = \frac{\mathbf{P}_{\mathbf{v}}^{\perp}\mathbf{a}_{cc}(t)}{\|\mathbf{P}_{\mathbf{v}}^{\perp}\mathbf{a}_{cc}(t)\|} \tag{35}$$

The subspace spanned by $\mathbf{u}_1$ and $\mathbf{u}_2$ is called the osculating plane. It represents the plane that best fits the manifold curve at the current moment. The osculating plane effectively reveals the geometric structure information of the array manifold.

$$\text{Span}\{\mathbf{u}_1, \mathbf{u}_2\} \equiv \text{Span}\{\mathbf{v}(t), \mathbf{a}_{cc}(t)\} \tag{36}$$

When the manifold is not confined to a two-dimensional plane, a third coordinate axis—the binormal vector $\mathbf{u}_3$—needs to be introduced. In the $l_{arc}$-domain, $\mathbf{u}_3$ is defined as the unit vector perpendicular to the osculating plane.

$$\mathbf{u}_3(l_{arc}) \,\|\, \left(\frac{d\mathbf{u}_2}{dl_{arc}} + \kappa_1\mathbf{u}_1\right) \tag{37}$$

Since linear combinations of $\mathbf{x}, \mathbf{v}, \mathbf{a}_{cc}$ all lie within the osculating plane, only the high-order dynamic terms contained in $\mathbf{j}$ can possibly point out of the plane in the $t$-domain. Define the orthogonal projection operator onto the osculating plane as $\mathbf{P}_{\text{osc}}^{\perp} = \mathbf{I} - \mathbf{u}_1\mathbf{u}_1^H - \mathbf{u}_2\mathbf{u}_2^H$. Using projection coefficient

$$d_1 = \langle \mathbf{u}_1, \mathbf{j} \rangle_{\mathbb{R}} = \text{Re}(\mathbf{u}_1^H\mathbf{j}) = -\frac{\mathbf{x}^H(\mathbf{\Omega}\ddot{\mathbf{\Omega}} + \mathbf{\Omega}^4)\mathbf{x}}{\|\mathbf{\Omega x}\|} \tag{38}$$

$$d_2 = \langle \mathbf{u}_2, \mathbf{j} \rangle_{\mathbb{R}} = \text{Re}(\mathbf{u}_2^H\mathbf{j}) = \frac{\mathbf{x}^H(2\mathbf{\Omega}^3\dot{\mathbf{\Omega}} - \dot{\mathbf{\Omega}}\ddot{\mathbf{\Omega}})\mathbf{x} - c_1d_1}{\left\|(\dot{\mathbf{\Omega}} + \mathbf{\Omega}^2)\mathbf{x} - c_1\dfrac{\mathbf{\Omega x}}{\|\mathbf{\Omega x}\|}\right\|} \tag{39}$$

The binormal vector can be expressed as

$$\mathbf{P}_{\text{osc}}^{\perp}\mathbf{j}(t) = (\mathbf{\Omega}^3 + 3\mathbf{\Omega}\dot{\mathbf{\Omega}} + \ddot{\mathbf{\Omega}})\mathbf{x} - d_1\mathbf{u}_1 - d_2\mathbf{u}_2 \tag{40}$$

$$\mathbf{u}_3(t) = \frac{\mathbf{P}_{\text{osc}}^{\perp}\mathbf{j}(t)}{\|\mathbf{P}_{\text{osc}}^{\perp}\mathbf{j}(t)\|} \tag{41}$$

Through the above derivations, we have established a mapping between physical derivatives and the geometric frame, which indicates that the high-order time derivatives of the manfoid are physical probes for detecting the high-dimensional geometric shape of the manifold.

*D. The Curvature and The Torsion*

This section derives the analytical expressions for the curvature $\kappa_1$ and torsion $\kappa_2$ of the manifold. We will demonstrate how to calculate these geometric invariants using orthogonal projections of physical derivatives in the real embedding space, and reveal their mapping relationships with signal dynamic parameters and the array geometric structure.

According to the Frenet-Serret frame, curvature is

$$\kappa_1(l_{arc}) = \|d\mathbf{u}_1 / dl_{arc}\| \tag{42}$$

Using the metric conversion relationship $dl_{arc}/dt = \|\mathbf{v}(t)\|$, we transform this into a time-domain projection formula based on the physical acceleration $\mathbf{a}_{cc}(t)$

$$\kappa_1(t) = \frac{\|\mathbf{P}_{\mathbf{v}}^{\perp}\mathbf{a}_{cc}(t)\|}{\|\mathbf{v}(t)\|^2} \tag{43}$$

Within the framework of real Riemannian geometry, utilizing the anti-Hermitian property of $\mathbf{\Omega}$, we obtain the closed-form solution for the first-order curvature

$$\kappa_1(t) = \frac{(\dot{\mathbf{\Omega}} + \mathbf{\Omega}^2)\mathbf{x} - c_1\dfrac{\mathbf{\Omega x}}{\|\mathbf{\Omega x}\|}}{\|\mathbf{\Omega x}\|^2} \tag{44}$$

Furthermore, we can obtain an approximate analytical expression (see Appendix A for details).

$$\kappa_1(\theta,t) \approx \sqrt{\frac{1}{M} + \left(\frac{2}{\sqrt{M}}\frac{\dot{\omega}(t)}{\omega(t)}\mathrm{std}(\boldsymbol{\tau}(\theta))\right)^2} = \sqrt{\kappa_{\mathrm{geo}}^2 + \kappa_{dyn}^2} \tag{45}$$

This formula reveals the orthogonal synthesis law of curvature: $\kappa_{\mathrm{geo}} = 1/\sqrt{M}$ is the curvature of the geodesic, originates from the hypersphere of radius $\sqrt{M}$ . Dynamic curvature $\kappa_{dyn}$ originates from the stretching of the manifold by the signal dynamics $\dot{\omega}$ , varies monotonically with the variance of the DOA distribution $\mathrm{std}(\boldsymbol{\tau})$ , constituting the geometric basis for DOA estimation.

The second-order torsion describes the degree to which the manifold trajectory deviates from the osculating plane.

$$\kappa_2(l_{arc}) = \| \frac{d\mathbf{u}_2}{dl_{arc}} + \kappa_1 \mathbf{u}_1 \| \tag{46}$$

As shown in (36), since $\mathbf{v}$ and $\mathbf{a}_{cc}$ span the osculating plane, torsion is entirely determined by the component of $\mathbf{j}(t)$ that lies outside this plane. The torsion calculation formula is

$$\kappa_2(t) = \frac{\left\|\mathbf{P}_{\mathrm{osc}}^{\perp}\mathbf{j}(t)\right\|}{\kappa_1(t)\left\|\mathbf{v}(t)\right\|^3} \tag{47}$$

Substituting the operator expansion into the formula.

$$\kappa_2(t) = \frac{\left\|(\mathbf{\Omega}^3 + 3\mathbf{\Omega}\dot{\mathbf{\Omega}} + \ddot{\mathbf{\Omega}})\mathbf{x} - d_1\mathbf{u}_1 - d_2\mathbf{u}_2\right\|}{\left\|\mathbf{\Omega}^2\mathbf{x}\right\| \cdot \left\|\mathbf{\Omega}\mathbf{x}\right\|} \tag{48}$$

We obtain the approximate analytical law for torsion (derivation provided in Appendix B).

$$\kappa_2(\theta,t) \approx \frac{2}{\sqrt{M}}\left|\frac{\dot{\omega}(t)}{\omega(t)}\right| \mathrm{std}(\boldsymbol{\tau}(\theta)) = \left|\kappa_{dyn}\right| \tag{49}$$

In summary, this work is not merely an extension of traditional array processing, but a complete geometric system. However, the analytical expressions for higher-order curvatures are complex. The analytical expressions for $\kappa_1(t)$ and $\kappa_2(t)$ derived in this paper also neglect the higher-order differential dynamics of the signal. Structural errors exist under certain parameters.

### *E. High-Order Extension*

This section extends the dynamic manifold framework to high-order geometric features, constructing a complete theoretical description framework.

From the perspective of theoretical completeness, a smooth curve embedded in an $M$ -dimensional complex space is uniquely described by the Frenet-Serret Frame, which consists of $2M$ generalized orthogonal unit basis vectors $\{\mathbf{u}_1,\ldots,\mathbf{u}_{2M}\}$ and $2M-1$ generalized curvatures $\{\kappa_1,\ldots,\kappa_{2M-1}\}$ .

The rotational motion of the manifold frame with respect to the parameter $l_{arc}$ follows a system of differential equations, whose structure is described by the Cartan Matrix $\mathbf{C}(l_{arc})$ [22].

TABLE I
DYNAMICS-GEOMETRY DUALITY HIERARCHY

| Geometric Object | Geometric parameters | Related physical parameter |
|---|---|---|
| Manifold | $\mathbf{x}(l_{arc})$ | $\frac{dl_{arc}}{dt} = \|\mathbf{v}(t)\|$ |
| Tangent vector | $\mathbf{u}_1(l_{arc}) = \frac{d\mathbf{x}}{dl_{arc}}$ | $\mathbf{u}_1(t) = \frac{\mathbf{v}(t)}{\|\mathbf{v}(t)\|}$ |
| Principal Normal Vector | $\mathbf{u}_2(l_{arc}) = \frac{\mathbf{u}_1'}{\|\mathbf{u}_1'\|}$ | $\mathbf{u}_2(t) = \frac{\mathbf{a}_\perp(t)}{\|\mathbf{a}_\perp(t)\|}$ |
| Binormal Vector | $\mathbf{u}_3(l_{arc}) = \frac{\mathbf{u}_2' + \kappa_1\mathbf{u}_1}{\|\mathbf{u}_2' + \kappa_1\mathbf{u}_1\|}$ | $\mathbf{u}_3(t) = \frac{\mathbf{j}_\perp(t)}{\|\mathbf{j}_\perp(t)\|}$ |
| Unit Basis Vector | $\mathbf{u}_i = \frac{\mathbf{u}_{i-1}' + \kappa_{i-2}\mathbf{u}_{i-2}}{\kappa_{i-1}}$ | $\mathbf{u}_i = \frac{\mathbf{x}_\perp^{(i)}(t)}{\|\mathbf{x}_\perp^{(i)}(t)\|}$ |
| Curvature | $\kappa_1 = \left\|\frac{d\mathbf{u}_1(l_{arc})}{dl_{arc}}\right\|$ | $\kappa_1 = \frac{\|\mathbf{a}_\perp(t)\|}{\|\mathbf{v}(t)\|^2}$ |
| Torsion | $\kappa_2 = \left\|\frac{d\mathbf{u}_2}{dl_{arc}} + \kappa_1\mathbf{u}_1\right\|$ | $\kappa_2 = \frac{\|\mathbf{j}_\perp(t)\|}{\kappa_1(t)\|\mathbf{v}(t)\|^3}$ |
| High-order Curvature | $\kappa_i = \|\mathbf{u}_i' + \kappa_{i-1}\mathbf{u}_{i-1}\|$ | $\kappa_i = \frac{\|\mathbf{x}_\perp^{(i+1)}\|}{\kappa_1\cdots\kappa_{i-1}\|\mathbf{v}\|^{i+1}}$ |

$$\frac{d}{dl_{arc}}\begin{bmatrix}\mathbf{u}_1\\ \mathbf{u}_2\\ \vdots\\ \mathbf{u}_{2M}\end{bmatrix} = \underbrace{\begin{bmatrix}0 & \kappa_1 & 0 & \cdots\\ -\kappa_1 & 0 & \kappa_2 & \cdots\\ \vdots & \vdots & \vdots & \ddots\end{bmatrix}}_{\mathbf{C}(l_{arc})}\begin{bmatrix}\mathbf{u}_1\\ \mathbf{u}_2\\ \vdots\\ \mathbf{u}_{2M}\end{bmatrix} \tag{50}$$

This matrix form indicates that the high-order geometric shape of the manifold is successively defined by a series of high-order curvatures.

In summary, we can induce the Dynamics-Geometry duality hierarchy in wideband array signal processing. The time-domain differential properties of the signal map directly to the geometric properties. This correspondence is summarized in Table I, where $(\cdot)_\perp$ denote the projection operator. This hierarchy table not only summarizes the theoretical findings of this paper but also points out the theoretical potential of using high-order dynamic information for finer signal analysis.

## IV. GEOMETRIC INVARIANCE AND ESTIMATION FRAMEWORKS

### *A. Dynamic Subspace Expansion*

This section aims to establish a new framework for unambiguous direction finding based on dynamic manifolds. To understand the physical basis of this framework, we need to examine the complete signal space.

Consider the high-order differential structure of the observation vector $\mathbf{x}(t)$ . According to the analysis in Chapter 3, the derivative $\mathbf{x}^{(k)}$ of any order $k$ is recursively generated by the dynamic generator $\mathbf{\Omega}$ and its time derivatives.

Consequently, the local evolution trajectory of the manifold at time $t$ is confined within a linear subspace generated by $\mathbf{\Omega}(t)$

$$\begin{aligned} \mathcal{S}_{dyn}(t) &= \text{span}\left\{\mathbf{x}(t), \mathbf{v}(t), \mathbf{a}_{cc}(t), \ldots\right\} \\ &\equiv \text{span}\left\{\mathbf{x}, \mathbf{\Omega x}, (\mathbf{\Omega}^2 + \dot{\mathbf{\Omega}})\mathbf{x}, \ldots\right\} \end{aligned} \tag{51}$$

In numerical linear algebra, this space can be viewed as the Generalized Krylov Subspace generated by the dynamic generator $\mathbf{\Omega}(t)$. From the perspective of differential geometry, the generalized orthogonal basis of this subspace at time $t$ is precisely the Frenet-Serret Frame $\{\mathbf{u}_1, \mathbf{u}_2, \ldots\}$.

The dimension of this subspace determines the observable degrees of freedom of the system. In traditional array signal processing, the signal space is spanned solely by $\mathbf{x}$ (Rank-1). In dynamic manifold, we are observing this high-dimensional expanded subspace $\mathcal{S}_{dyn}$. The DOA acts on the array topology, introducing the dynamic evolution directions of the manifold represented by the high-order terms. Therefore, the essence of dynamic direction finding is utilizing the expanded degrees of freedom provided by the generalized Krylov subspace to extract geometric attributes of the manifold.

Based on the subspace definition above, we can fundamentally explain how the dynamic manifold breaks the limitations of the spatial sampling theorem to achieve unambiguity direction finding.

Assume that in a sparse array, there exist two different directions $\theta_1, \theta_2$ whose steering vectors $\mathbf{a}(\theta)$ are linearly dependent (with coefficient $\alpha$), i.e.,

$$\mathbf{x}(\theta_1, t) = \alpha \cdot \mathbf{x}(\theta_2, t) \tag{52}$$

In the subspace $\text{span}\{\mathbf{x}\}$, these two directions are indistinguishable. We need to consider whether this same linear relationship is maintained in the defined $\mathcal{S}_{dyn}$ by test whether the dynamic evolution directions of any order or the derived geometric parameters maintain the same linear relationship. Examine the first-order term and define the residual vector as

$$\Delta\mathbf{v} = \mathbf{v}(\theta_1, t) - \alpha\mathbf{v}(\theta_2, t) . \tag{53}$$

Substituting the ambiguity condition $\mathbf{x}(\theta_1, t) = \alpha \cdot \mathbf{x}(\theta_2, t)$

$$\Delta\mathbf{v} = j\left[\sum_{n=1}^{\infty}\frac{(-1)^n}{n!}\omega^{(n)}(t)(\mathbf{T}^n(\theta_1) - \mathbf{T}^n(\theta_2))\right]\mathbf{x}(\theta_1, t) \tag{54}$$

Since $\mathbf{T}^n(\theta_1) \neq \mathbf{T}^n(\theta_2)$, as long as the first-order dynamic $\dot{\omega}(t)$ is not constantly zero, the tangent vector no longer satisfies the linear dependence relationship. Consequently, the augmented vectors in the high-dimensional expanded subspace $\mathcal{S}_{dyn}$ are linearly independent. The dynamic generator $\mathbf{\Omega}$ destroys the linear correlation in the static space by leveraging its structural differences.

Based on above analysis, we propose the general theorem.

**Theorem 1**. The dynamic manifold of a signal is a curve in the high-dimensional observation space. The DOA $\theta$ is an intrinsic structural parameter encoded in the curve's differential geometric features independent of spatial aliasing.

The validity of this theorem strictly depends on whether the subspace is successfully expanded, i.e., whether the aforementioned linear independence holds. This is defined by the following two conditions:

1. If $\dot{\omega}$ is identically zero (e.g., Monopulse (MP) signal), the operator degenerates to scalar multiplication $\mathbf{\Omega} = j\omega\mathbf{I}$. In this case, the $\Delta\mathbf{T}$ term vanishes, causing frames of any order to remain linearly dependent. The system degenerates to traditional array signal processing.
2. The array geometric structure must be non-degenerate, meaning the delay matrix corresponding to different angles must be separable $\mathbf{T}_1 \neq \mathbf{T}_2$.

In summary, the theorem 1 extends the direction-finding capabilities of sparse arrays and provides a theoretical foundation for designing new wideband array processing algorithms.

### *B. Geometric Invariance Theory and Robust Estimator*

We have established the feasibility of unambiguous direction finding using dynamic manifolds in theorem 1. Additionally, the manifold possesses numerous geometric attributes that exhibit unique properties. Among these, the most prominent is rigid body transformation invariance.

In practical systems, unknown phase error is mathematically equivalent to applying a rigid body transformation to the observation manifold.

$$\tilde{\mathbf{x}}(t) = \mathbf{\Gamma x}(t) \tag{55}$$

Let $\mathcal{F}_{geo}(\cdot)$ denote a differential mapping of manifold parameters. If a geometric parameter remains unchanged under rigid body transformation, i.e., $\mathcal{F}_{geo}(\tilde{\mathbf{x}}) = \mathcal{F}_{geo}(\mathbf{x})$, we classify it as an intrinsic invariant of the dynamic manifold. Parameters that do not satisfy this condition are termed extrinsic parameter.

According to the analysis in Chapter 3, the derivative $\mathbf{x}^{(k)}$ is recursively generated by the dynamic generator $\mathbf{\Omega}$

$$\mathcal{F}_{geo}(\mathbf{x}) = \mathbf{x}^{(k)} = \mathcal{F}_k(\mathbf{\Omega})\mathbf{x} . \tag{56}$$

Therefore, vector features of all orders (such as $\mathbf{v}, \mathbf{a}, \mathbf{u}_i$) will rotate with the rigid body transformation and thus belong to the extrinsic parameters

$$\mathcal{F}_{geo}(\tilde{\mathbf{x}}) = \mathcal{F}_k(\mathbf{\Omega})\tilde{\mathbf{x}} = \mathbf{\Gamma} \cdot \mathcal{F}_{geo}(\mathbf{x}) \tag{57}$$

On the contrary, the shape of a curve (scalars such as curvature and torsion) is an invariant of motion in Euclidean space. Examining the curvature

$$\mathcal{F}_{geo}(\tilde{\mathbf{x}}) = \tilde{\kappa}_1(t) = \frac{\left\|\mathbf{P}_{\mathbf{\Gamma v}}^{\perp}(\mathbf{\Gamma a})\right\|}{\left\|\mathbf{\Gamma v}\right\|^2} = \frac{\left\|\mathbf{\Gamma}(\mathbf{P}_{\mathbf{v}}^{\perp}\mathbf{a})\right\|}{\left\|\mathbf{v}\right\|^2} = \kappa_1(t) = \mathcal{F}_{geo}(\mathbf{x}) \tag{58}$$

This implies that the shape of the curve, i.e., curvatures of all orders, belongs to intrinsic invariant. Based on the behavior of geometric features under rigid body transformation, we introduce a new theorem.

**Theorem 2**. The extrinsic parameters of the dynamic manifold contain the attitude information of the manifold; they are not limited by the spatial sampling theorem but are constrained by rigid body transformations caused by array phase errors. The intrinsic invariants describe only the inherent shape information of the manifold, and thus naturally possess robustness against both the spatial sampling theorem limits and channel errors.

Based on Theorem 2, we propose two feasible generalized estimator frameworks utilizing geometric parameters for DOA estimation.

Framework I is based on extrinsic vector features ( $\mathbf{v}, \mathbf{a}, \mathbf{u}_i$ ). Its geometric essence is searching for a theoretical manifold $\mathcal{M}(\theta)$ in the phase space that maximizes the overlap (via inner product or projection) with the observed trajectory. For an example, we define a general coherent cost functional.

$$\hat{\theta} = \arg\max_{\theta} \int_{T_{obs}} \mathcal{C}_{coh}\left(\mathbf{v}_{obs}(t), \mathbf{v}_{model}(\theta, t)\right) dt \tag{59}$$

where $\mathcal{C}_{coh}(\cdot)$ is a coherent metric operator. This estimator is essentially a direct generalization of traditional array signal processing within the dynamic manifold framework. However, if an unknown phase error $\mathbf{\Gamma}$ exists, the extrinsic vector parameters rotate, causing ambiguity with the model.

Framework II is based on the intrinsic invariants ( $\kappa$ ). This is the core contribution of this paper distinguishing it from traditional array processing. Its geometric essence is to abandon the matching of manifold position and direction, turning instead to matching the topological shape of the manifold. For an example, we define a general feature matching cost function.

$$\hat{\theta} = \arg\min_{\theta} \int_{T_{obs}} \left\| \boldsymbol{\kappa}_{obs}(t) - \boldsymbol{\kappa}_{model}(\theta, t) \right\|^2 dt \tag{60}$$

where $\boldsymbol{\kappa} = [\kappa_1, \kappa_2, \ldots]^T$ is the vector of geometric invariants. According to Theorem 2, even in the presence of unknown channel phase errors $\mathbf{\Gamma}$, the observed curvature $\kappa_{obs}$ remains equal to the theoretical curvature $\kappa_{model}$ . Therefore, this framework enables calibration-free direction finding.

However, it must be noted that feature extraction involves high-order differential operations, which amplifies noise. Therefore, this framework is more suitable for high signal-to-noise ratio (SNR) environments or large aperture array.

## V. Simulation Experiments and Analysis

This chapter conducts a validation of the dynamic manifold framework through numerical simulations. Unless otherwise specified, the simulation sampling rate is set to satisfy the requirements for high-precision differentiation. Three typical signal types are simulated: a MP signal, a LFM signal with a bandwidth of 800 MHz, and a SFM signal with a modulation slope of 10 MHz and a modulation index of 15. All signals share a pulse width of 200 ns and a carrier frequency of 2 GHz.

First experiment aims to verify the core hypothesis that the time-domain dynamic characteristics of a signal directly determine the geometric topology of the dynamic manifold, as shown in Fig. 3. We consider a three-element sparse linear array with an element spacing $[0,5,10]d$ , where $d$ representing a half wavelength. Under a SNR of 30 dB, the three types of signals are generated, and the trajectories of the observation vectors are extracted via Principal Component Analysis (PCA) for three-dimensional visualization.

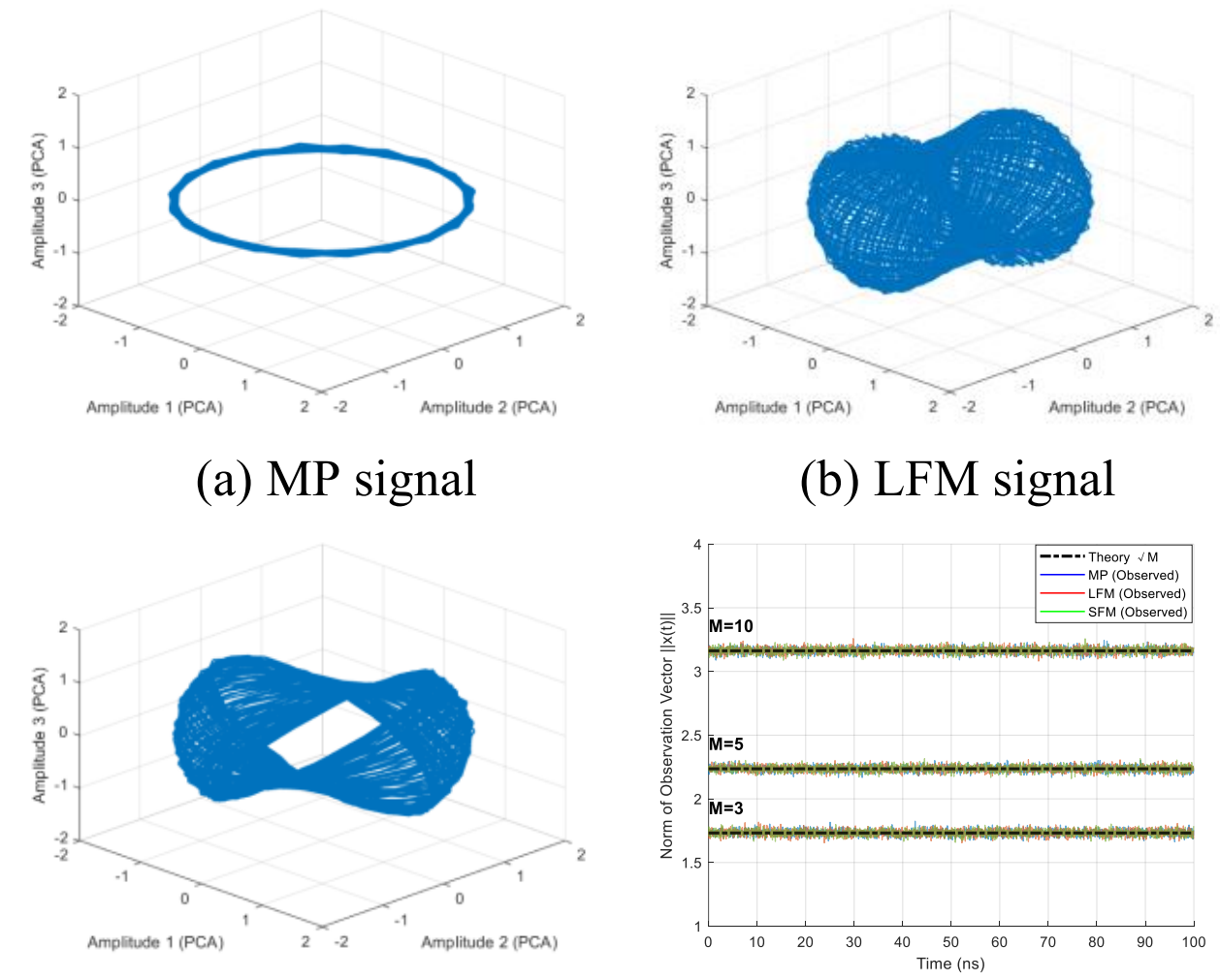

(a) MP signal (b) LFM signal

(c) SFM signal (d) The observation norm

**Fig. 3.** The geometric topology of the observation manifold

The trajectory of the MP signal closes into a closed circle, as shown in Fig. 3(a), where the manifold is constrained solely by the constant baseline geodesic curvature. In contrast, the LFM signal breaks this closed loop; as depicted in Fig. 3(b), its trajectory extends along the binormal direction. The SFM signal exhibits an even more complex spatially twisted curve, as shown in Fig. 3(c). Furthermore, the numerical values of the observation vector norm $\|\mathbf{x}\|$ for different numbers of array elements are presented in Fig. 3(d). These results verify the constraint that the signal manifold is strictly confined to a high-dimensional hypersphere with a radius of $\sqrt{M}$ .

The simulation results intuitively confirm that as the order of signal dynamics increases, the manifold evolves from a two-dimensional closed curve into a three-dimensional spatial curve. This expansion of geometric dimensionality represents the pivotal mechanism by which the dynamic manifold framework resolves traditional array ambiguities.

To verify whether the dynamic generator derived in Chapter 3 can accurately describe the high-order evolution of the manifold, the second experiment designs a comparative test between theoretical values and measured values. To exclude the masking of waveform details by strong noise and focus on verifying the correctness of the geometric structure, the SNR is set to 40dB. The array configuration remains consistent with Experiment 1, and we examine two representative dynamic signals: LFM and SFM. The physical measurement curves are obtained by directly applying a high-order Savitzky-Golay filter to the noisy observation data for numerical differentiation, successively extracting velocity $\mathbf{v}$ , acceleration $\mathbf{a}$ , and jerk $\mathbf{j}$ . The theoretical truth curves are analytically generated based on the dynamic operator $\mathbf{\Omega}$ using the chain rule. The experimental results are shown in Fig. 4.

Whether for LFM or SFM signals, the theoretical curves and the physical measurement curves coincide highly in both trend and numerical value. This confirms the mathematical correctness and its high-order generalizations.

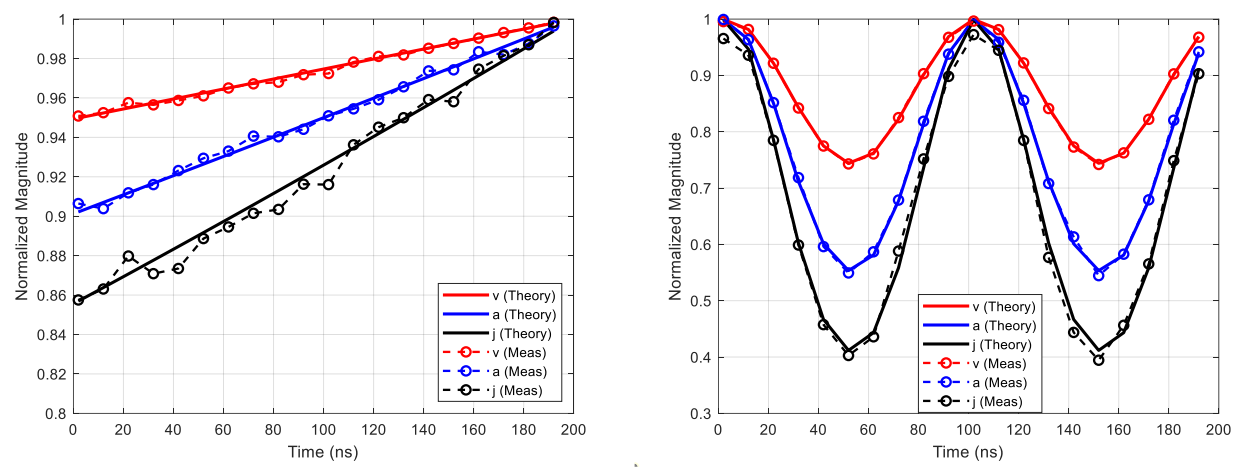

(a) LFM signal (b) SFM signal

**Fig. 4.** Comparison of numerical differentiation and theoretical formulas

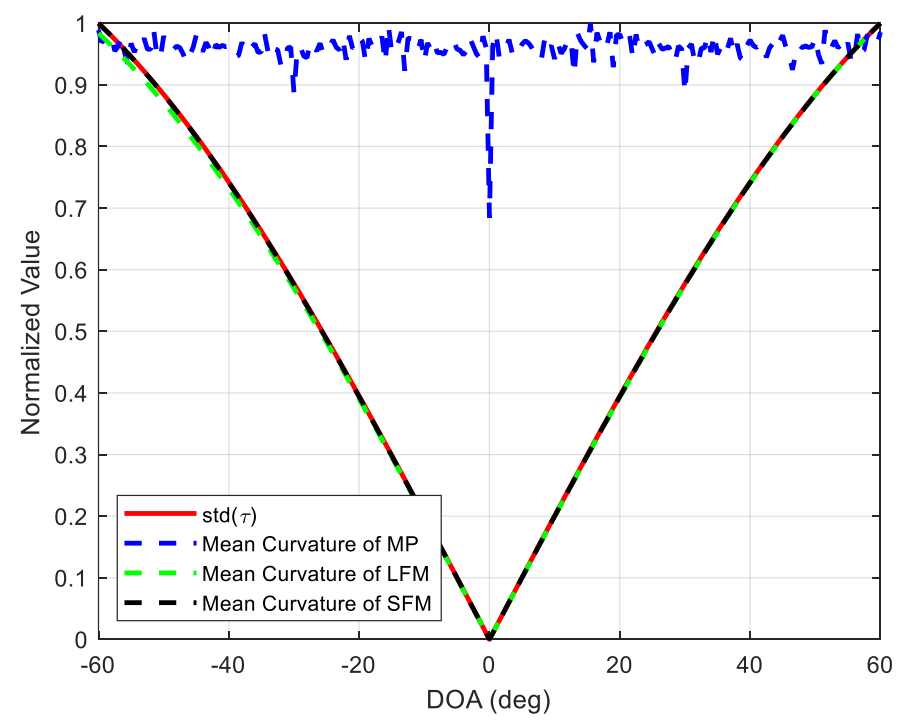

**Fig. 5.** Comparison of average curvatures

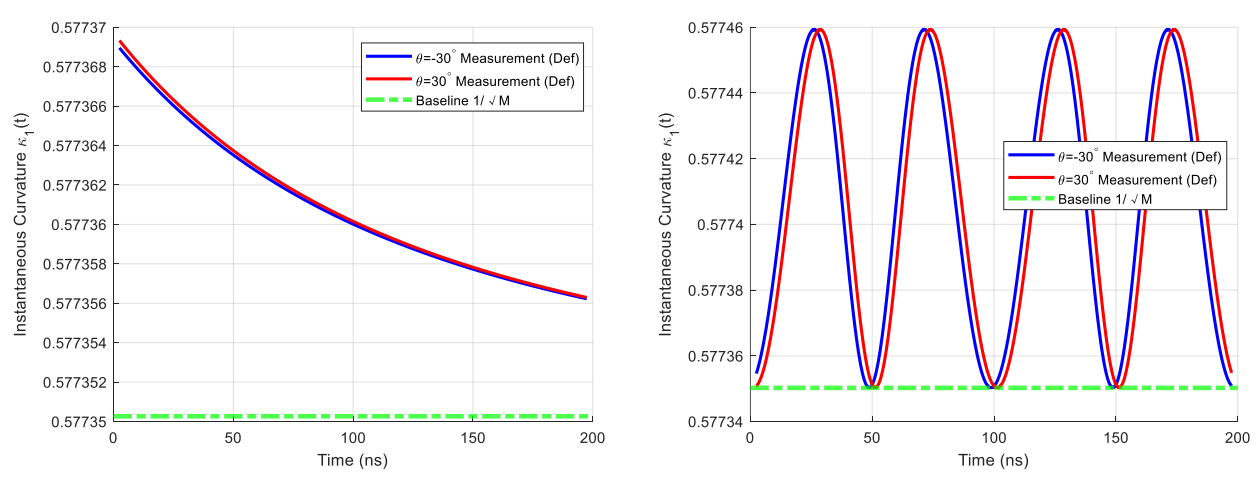

(a) LFM signal (b) SFM signal

**Fig. 6.** Instantaneous curvature under symmetric angles for a non-centered array

Next, array parameters changed into $[-5,0,5]d$ , which is a centralized array. Fig. 5 illustrates the trend of the average curvature of three signal types with respect to the array geometric factor $\mathrm{std}(\boldsymbol{\tau})$ . The results indicate that the curvature of the MP signal exhibits noise-like characteristics, suggesting that its geometric features do not vary with DOA. Both LFM and SFM signals show a high degree of linear correlation with $\mathrm{std}(\boldsymbol{\tau})$ . This confirms that the dynamic perturbation term $\kappa_{dyn}$ linearly encodes the DOA into the bending degree of the manifold, providing a physical basis for geometric direction finding. For non-centralized arrays, positive and negative DOA's curvature possess distinctiveness.

Fig. 6 compares the curvature under symmetric angles ( $\pm 30^{\circ}$ ) for a non-centered array $[0,5,10]d$ . For the LFM signal, as shown in Fig. 6(a), the overall curvature amplitude at $+30^{\circ}$ is higher than at $-30^{\circ}$ . Similarly, the SFM signal, as shown in Fig. 6(b), shows a phase dislocation between peaks and troughs under positive and negative angles. This manifold chirality (asymmetry) induced by the Geometric Doppler Effect equips the dynamic manifold method with the capability to resolve the sign of the DOA.

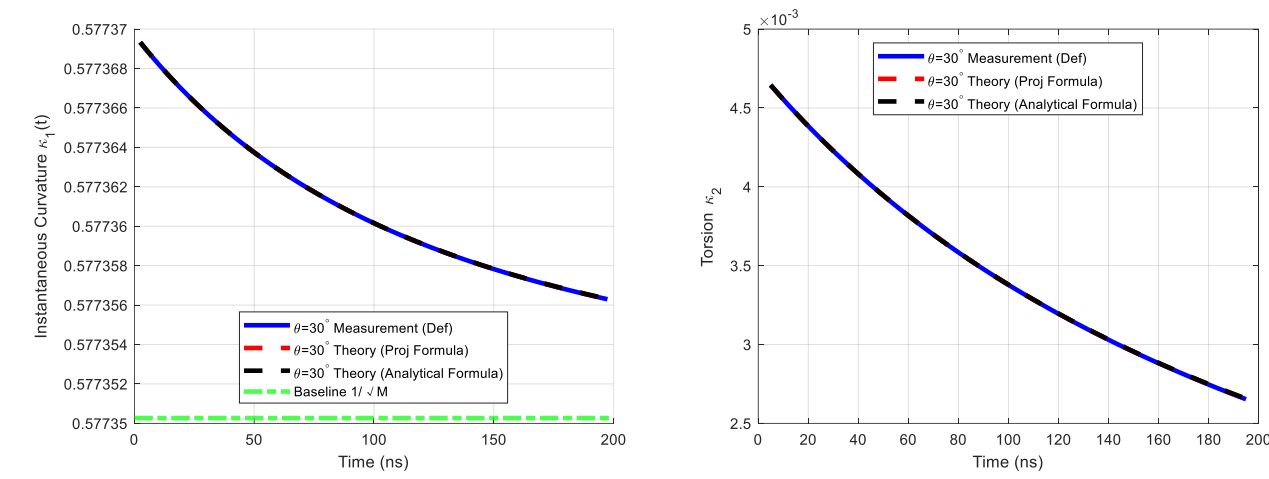

(a) Curvature of small aperture(b) Torsion of small aperture

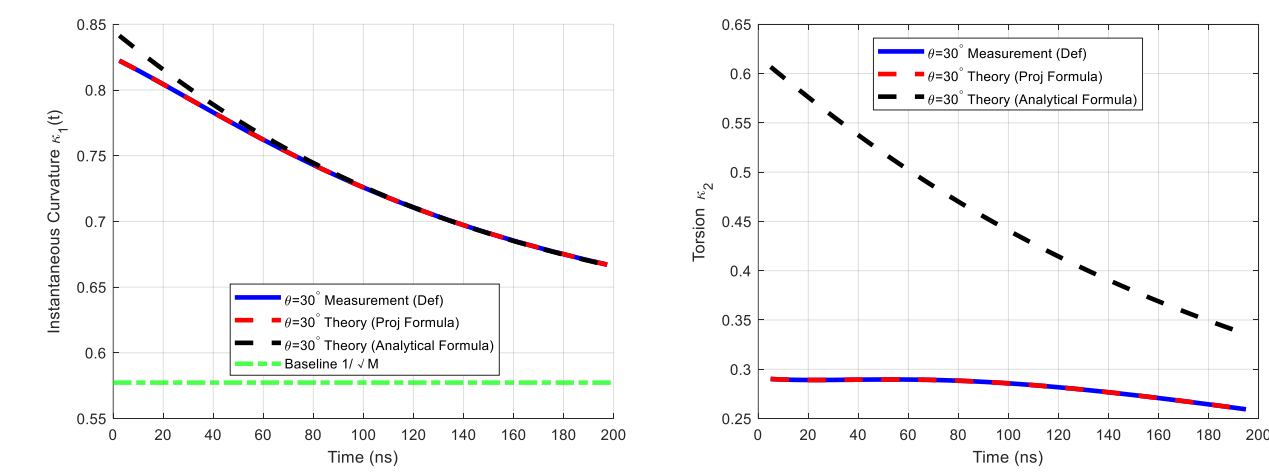

(c) Curvature of large aperture (d) Torsion of large aperture

**Fig. 7.** Curvature and torsion of LFM signal

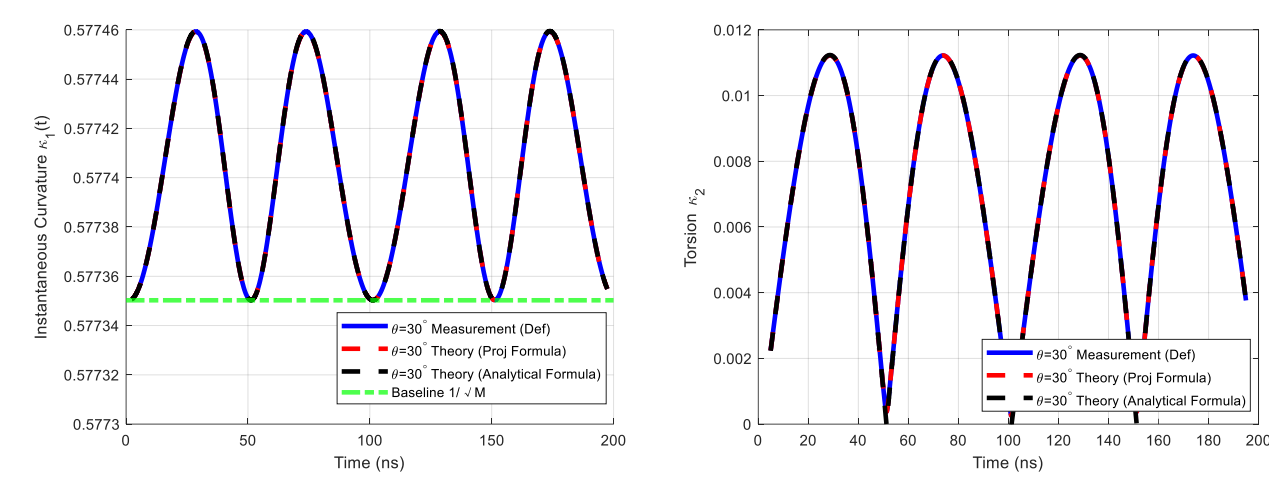

(a) Curvature of small aperture(b) Torsion of small aperture

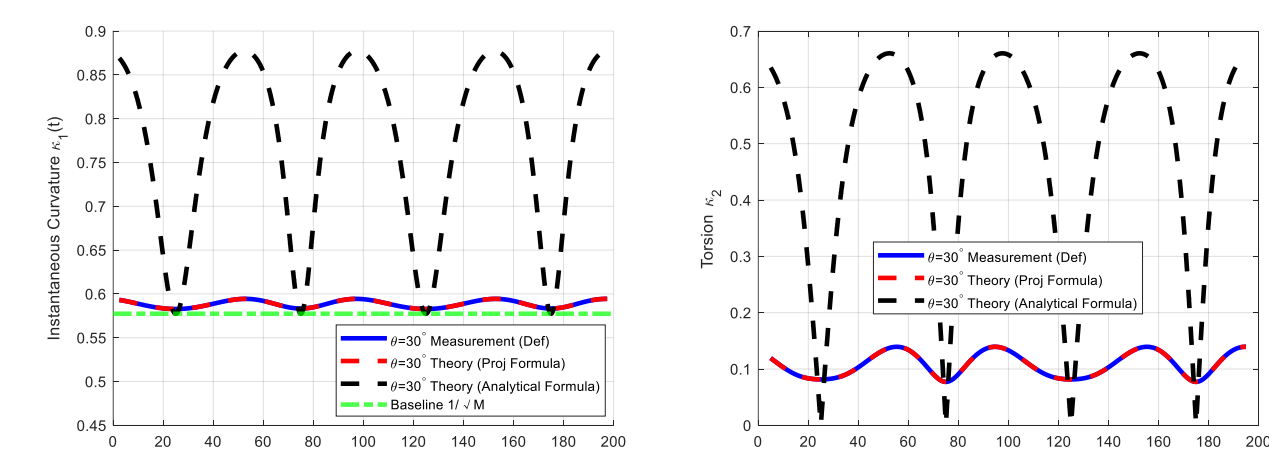

(c) Curvature of large aperture (d) Torsion of large aperture

**Fig. 8.** Curvature and torsion of SFM signal

What's more, we examine the time-varying curves of geometric features for LFM (Fig. 7) and SFM (Fig. 8) under small aperture $[0,5,10]d$ and ultra-large aperture $[0,500,1000]d$ conditions, respectively. The experiment employs three methods: the geometric truth obtained by numerical differentiation of sampled data (Measurement); the precise projection result based on the dynamic operator $\boldsymbol{\Omega}$ (Projection); and the closed-form approximation obtained via perturbation expansion (Analytical).

Under small aperture conditions, as shown in Fig. 7(a) and Fig. 8(a), the curvature $\kappa_1$ for all signals clusters around the theoretical baseline value $1/\sqrt{M}$ (approximately 0.577). LFM appears as a monotonic curve, while SFM exhibits periodic oscillations. Simulation results show that the three curves overlap highly, validating the orthogonal synthesis law.

When the array spacing is expanded to $[0,500,1000]d$ , as shown in Fig. 7(c-d) and Fig. 8(c-d), the amplitudes of $\kappa_1$ and $\kappa_2$ increase. The term $\kappa_{dyn}$ is no longer a minute quantity but becomes a strong feature. This indicates that a large aperture amplifies geometric identifiability. However, under large aperture conditions, the algebraic approximation begins to show errors, whereas the projection remains precisely locked to the measurement. This demonstrates that under high delay conditions, the complete structure of $\mathbf{\Omega}$ must be considered.

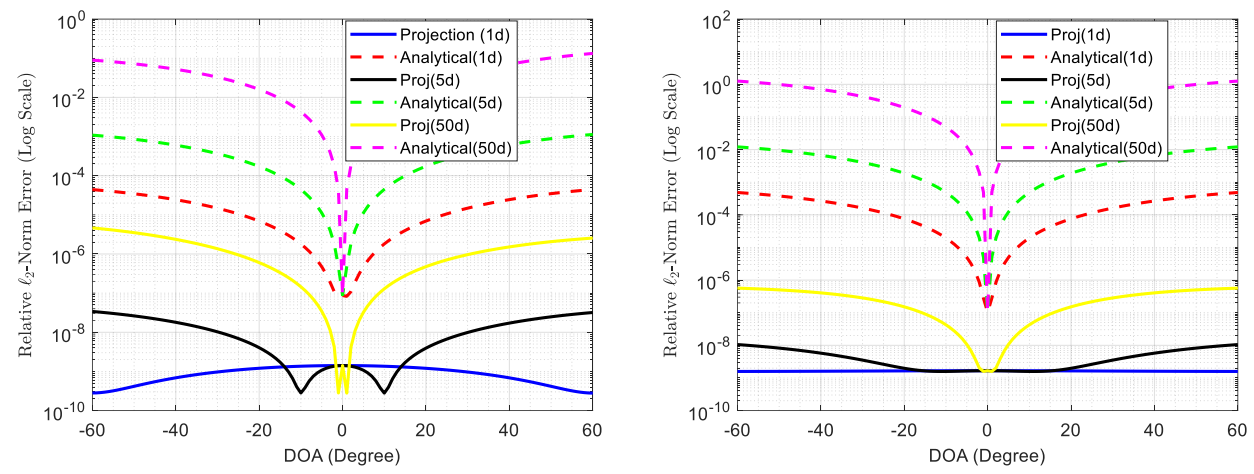

(a) LFM curvature (b) SFM torsion

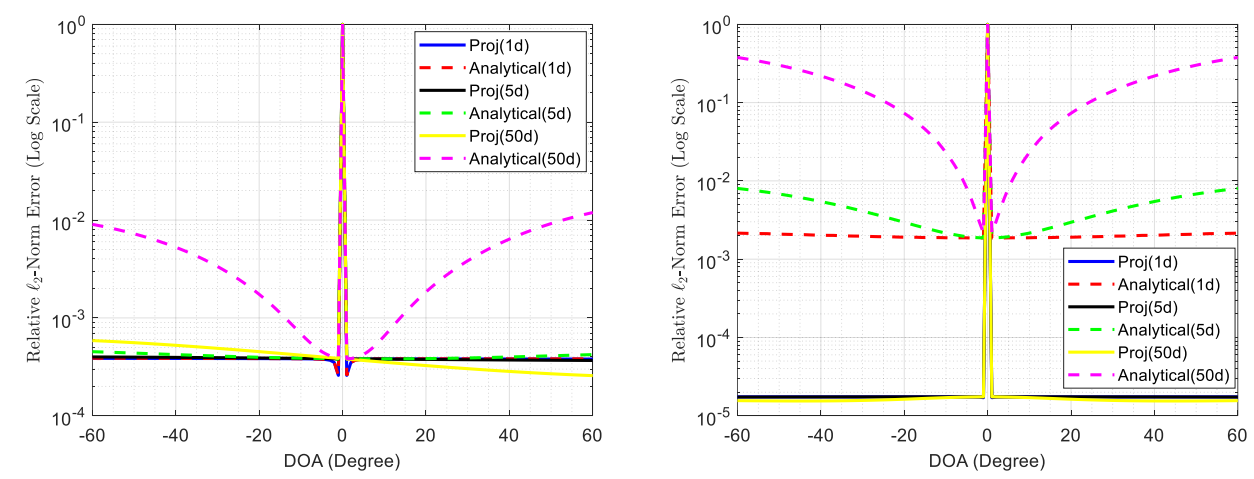

(a) SFM curvature (b) SFM torsion

**Fig. 9.** RMSE under different apertures and DOA

Fig. 9 quantifies the error laws of the two theoretical models relative to the physical truth as aperture and azimuth vary. The error is calculated as the $\ell_2$ -norm between the theoretical model and the geometric truth. Here, 1d represents a uniform linear array with spacing $[0,1,2]d$ , 5d represents $[0,5,10]d$ , and 50d represents $[0,50,100]d$ . The results show that the Projection Formula error remains consistently at an extremely low level, proving that the differential geometric description based on the dynamic operator $\mathbf{\Omega}$ is physically exact. The algebraic formula error rises with increasing aperture. This indicates that the closed-form approximation is valid only under small aperture conditions.

The next simulation verifies the correctness of the Framework I and the Framework II through multi-level simulation scenarios. The target signal DOA is set to $20^{\circ}$ .

First, we examine a MP signal incident on a uniform array with spacing 5d. The experimental results, as shown in Fig. 10, reveal that MUSIC exhibits 5 peaks of nearly equal amplitude. Since the dynamic features of MP is not available, the spectral peak morphology of Framework I is completely consistent with MUSIC, also showing 5 ambiguous peaks. This validates our theoretical inference that, when dynamics are absent, the dynamic subspace and Framework I degenerates into a conventional array processing method. The spectrum of Framework II approaches 0 or invalid values across the entire domain. This is because the MP signal manifold degenerates into a closed circle, where its geometric feature $\kappa(t)$ is constant and isotropic, losing the ability to encode the DOA.

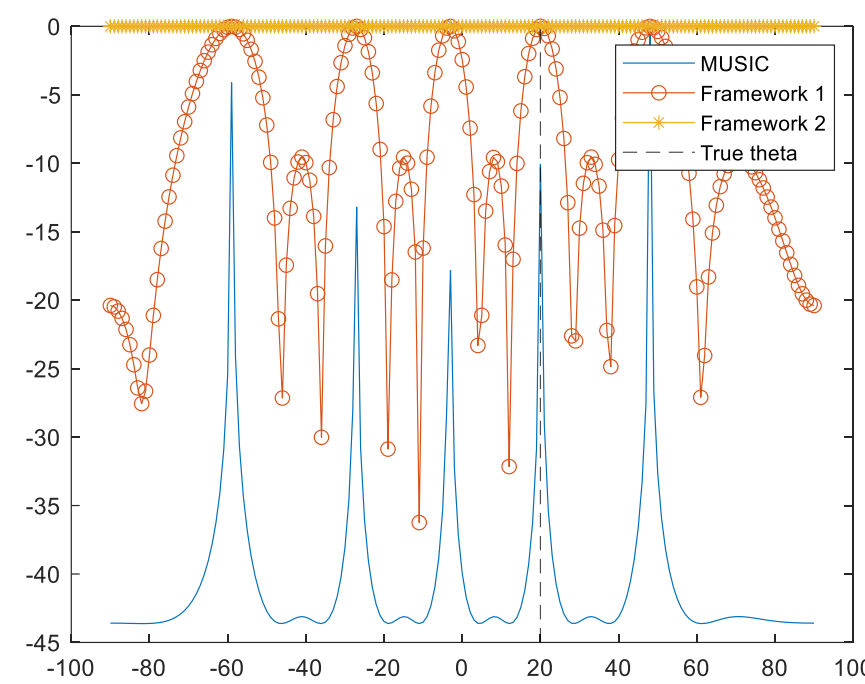

**Fig. 10.** DOA estimation results for a MP signal

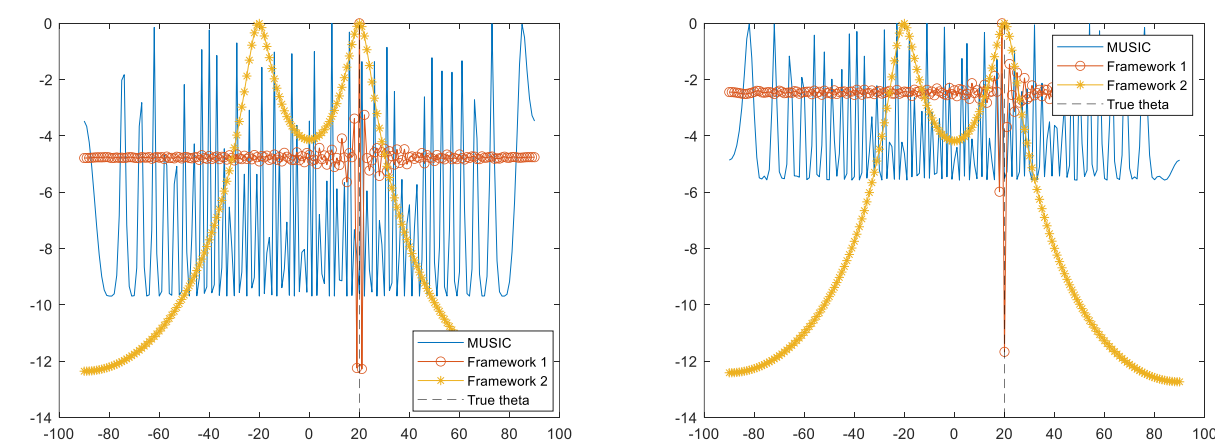

(a) No phase error (b) Have phase error

**Fig. 11.** Direction-finding results under small aperture

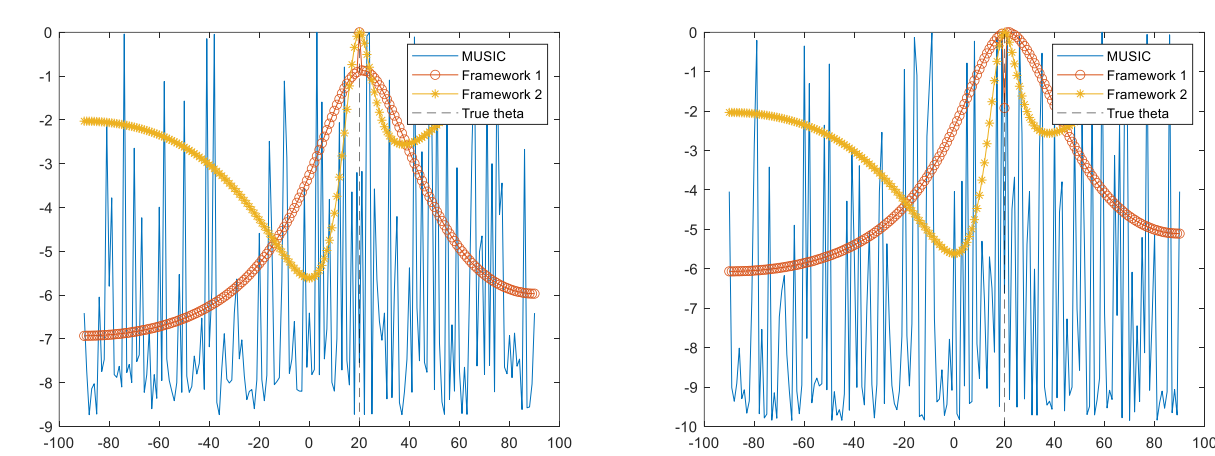

(a) No phase error (b) Have phase error

**Fig. 12.** Direction-finding results under large aperture

Next, an LFM signal is introduced under a 50d uniform array condition, utilizing a high SNR of 100 dB to highlight weak geometric perturbation features. The results are shown in Fig. 11(a). The MUSIC spectrum presents dense comb-like grating lobes. Framework I successfully suppresses all grating lobes, forming a single main peak pointing to $20^{\circ}$ . Framework II exhibits two symmetric main peaks of approximately equal amplitude, which can be distinguished by expanding the aperture or incorporating vector direction identification. This proves that whether using intrinsic or extrinsic parameters, dynamic information can break spatial ambiguity.

After introducing uncalibrated random element phase errors $\mathbf{\Gamma}$, as shown in Fig. 11(b), Framework I fails. Since phase errors cause a rigid body rotation misalignment between the theoretical model and the observation in the complex space, preventing coherent integration from forming an effective spectral peak. The result of Framework II remains unchanged.

To verify the impact of aperture on the frameworks, we expand the array aperture to 5000d (Ultra-Large Aperture) and reduce the SNR to 30 dB. The experimental results are shown in Fig. 12(a). Without phase error, both Framework I and II precisely lock onto the true value.

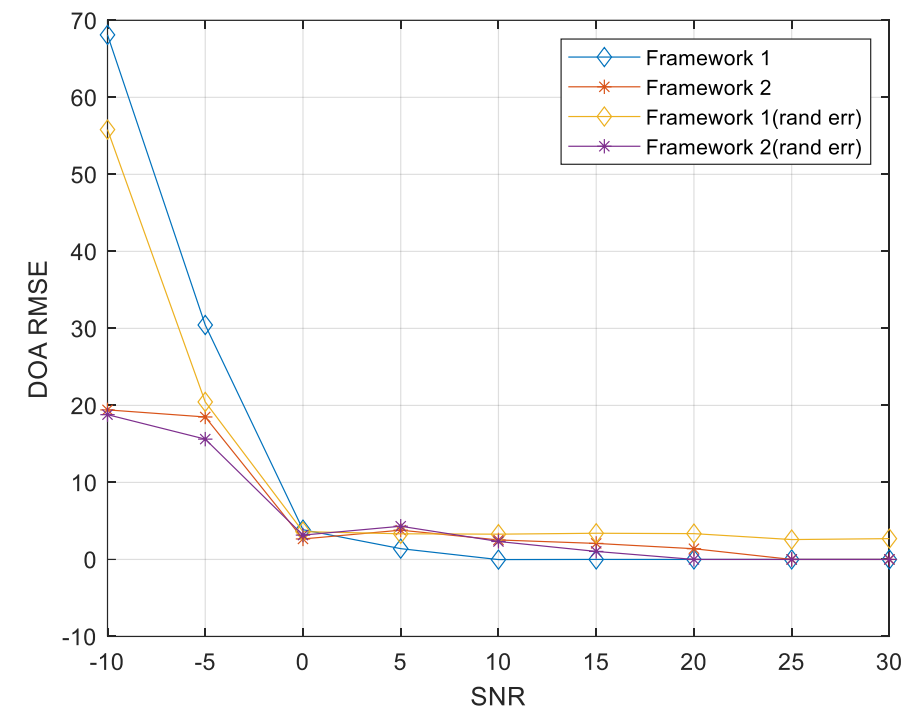

**Fig. 13.** RMSE of DOA estimation under different SNR for the proposed frameworks

With phase error, as shown in Fig. 12(b), Framework I shows a fixed bias of approximately $2^{\circ}$, while Framework II precisely points to the true DOA. This indicates that for Framework I, the rotation manifests as a systematic offset in beam pointing during large-aperture coherent integration. Conversely, the geometric invariance of Framework II holds strictly under large apertures, and due to the significant enhancement of geometric features, its noise immunity is also greatly improved.

Finally, to verify the statistical characteristics and applicable boundaries of the proposed theoretical frameworks, we conducted Monte Carlo simulations (100 independent trials) under a 5000d array to examine the variation of RMSE with SNR (-10 dB to 30 dB). The results are shown in Fig. 13. Framework I, in the absence of error, shows RMSE linearly decreasing with SNR, converging to 0; in the presence of error, RMSE exhibits a significant Error Floor, converging to $2.7^{\circ}$. Framework II, regardless of the presence of phase errors, shows RMSE continuously decreasing with improved SNR, finally converging to 0. This establishes the superiority of the intrinsic geometric framework in non-cooperative scenarios in a statistical sense.

## VI. Conclusion

This paper establishes a novel dynamic manifold perspective, offering a paradigm shift in UWB array signal processing from traditional phase consistency matching to geometric shape matching. By constructing a dual-parameter system comprising Extrinsic and Intrinsic parameters, we theoretically revealed how signal temporal dynamics translate into generalized identifiability capable of suppressing grating lobes, and how the geometric invariants of the manifold trajectory endow the system with intrinsic robustness against channel phase errors.

While this work lays a mathematical foundation and provides general estimator frameworks, specific algorithms remain to be developed. Future research should focus on designing robust and efficient direction-finding algorithms to fully exploit the expanded degrees of freedom offered by the dynamic manifold.

## Appendix

### Appendix A: First-Order Curvature

In Section 3.3, we presented the analytical formula for the first-order curvature. Here, we explicitly demonstrate the vector projection process.

In real Riemannian geometry, the projection coefficient is defined as $k_P = \frac{\langle \mathbf{v}, \mathbf{a}_{cc} \rangle_{\mathbb{R}}}{\|\mathbf{v}\|^2}$. To obtain an analytical solution with clear physical meaning, we need to evaluate the squared velocity magnitude in the denominator $\|\mathbf{v}(t)\|^2 = \sum_{m=1}^{M} \omega^2(t-\tau_m)$.

Without loss of generality, we set the delay reference point at the Geometric Centroid of the array, satisfying $\sum_{m=1}^{M} \tau_m(\theta) = 0$. Performing a second-order Taylor expansion on the squared instantaneous frequency yields

$$\sum_{m=1}^{M} \omega^2(t-\tau_m) \approx \sum_{m=1}^{M} \left[ \omega^2(t) - 2\omega(t)\dot{\omega}(t)\tau_m + \dot{\omega}^2(t)\tau_m^2 + O(\tau_m^n) \right] \tag{61}$$

Utilizing the centroid property $\sum \tau_m = 0$, the linear error term vanishes

$$\|\mathbf{v}(t)\|^2 \approx M\omega^2(t) + M\dot{\omega}^2(t)\mathrm{std}^2(\boldsymbol{\tau}) + \sum_{m=1}^{M} O(\tau_m^n) \tag{62}$$

For conventional radar and communication signals, the carrier frequency $\omega$ is much larger than the frequency variation caused by aperture transit time (i.e., $\omega \gg \dot{\omega}\tau_{max}$). Consequently, the contribution of higher-order terms to the magnitude is negligible, allowing for the high-precision approximation. Additionally, we can obtain that $\|\mathbf{x}\| = \sqrt{M}$, and

$$\|\mathbf{T}\mathbf{x}\| = \sqrt{\sum |\tau_m x_m|^2} = \sqrt{\sum \tau_m^2} = \sqrt{M} \cdot \mathrm{std}(\boldsymbol{\tau}). \tag{63}$$

Recall the first-order dynamic expansions for the tangent vector and acceleration vector

$$\mathbf{a}_{cc}(t) = \frac{d}{dt}\mathbf{v}(t) = \dot{\boldsymbol{\Omega}}\mathbf{x} + \boldsymbol{\Omega}\dot{\mathbf{x}} = (\dot{\boldsymbol{\Omega}} + \boldsymbol{\Omega}^2)\mathbf{x} \tag{64}$$

Substituting the operator dominant term approximation $\boldsymbol{\Omega} \approx j\omega\mathbf{I} - j\dot{\omega}\mathbf{T}$

$$\begin{aligned} &\dot{\boldsymbol{\Omega}} \approx j\dot{\omega}\mathbf{I} - j\ddot{\omega}\mathbf{T} \\ &\boldsymbol{\Omega}^2 \approx (j\omega\mathbf{I} - j\dot{\omega}\mathbf{T})^2 \approx -\omega^2\mathbf{I} + 2\omega\dot{\omega}\mathbf{T} - \dot{\omega}^2\mathbf{T}^2 \end{aligned} \tag{65}$$

We can obtain that the acceleration vector has five-term decomposition

$$\mathbf{a}_{cc} \approx \underset{a}{-\omega^2\mathbf{x}} + \underset{b}{j\dot{\omega}\mathbf{x}} - \underset{c}{j\ddot{\omega}\mathbf{T}\mathbf{x}} + \underbrace{2\omega\dot{\omega}\mathbf{T}\mathbf{x}}_{d} - \underbrace{\dot{\omega}^2\mathbf{T}^2\mathbf{x}}_{e} \tag{66}$$

It is needed to perform the real inner product operation between $\mathbf{v}$ and each of the five terms of $\mathbf{a}_{cc}$.

$$\begin{aligned} &\text{a.}\mathrm{Re}\left((-j\omega\mathbf{x}^H + j\dot{\omega}\mathbf{x}^H\mathbf{T})(-\omega^2\mathbf{x})\right) = 0 \\ &\text{b.}\mathrm{Re}\left((-j\omega\mathbf{x}^H + j\dot{\omega}\mathbf{x}^H\mathbf{T})(j\dot{\omega}\mathbf{x})\right) = M\omega\dot{\omega} \\ &\text{c.}\mathrm{Re}\left((-j\omega\mathbf{x}^H + j\dot{\omega}\mathbf{x}^H\mathbf{T})(-j\ddot{\omega}\mathbf{T}\mathbf{x})\right) = \dot{\omega}\ddot{\omega}M\mathrm{std}^2(\boldsymbol{\tau}) \\ &\text{d.}\mathrm{Re}\left((-j\omega\mathbf{x}^H + j\dot{\omega}\mathbf{x}^H\mathbf{T})(2\omega\dot{\omega}\mathbf{T}\mathbf{x})\right) = 0 \\ &\text{e.}\mathrm{Re}\left((-j\omega\mathbf{x}^H + j\dot{\omega}\mathbf{x}^H\mathbf{T})(-\dot{\omega}^2\mathbf{T}^2\mathbf{x})\right) = 0 \end{aligned} \tag{67}$$

The tangential projection coefficient $k_P$ is contributed by Term b and Term c

$$k=\frac{M\omega\dot{\omega}+\dot{\omega}\ddot{\omega}M\mathrm{std}^2(\boldsymbol{\tau})}{M\omega^2}=\frac{\dot{\omega}}{\omega}+\frac{\dot{\omega}\ddot{\omega}\mathrm{std}^2(\boldsymbol{\tau})}{\omega^2} \tag{68}$$

Calculate the tangential component vector

$$\begin{aligned}\mathbf{a}_{\parallel}&=k\mathbf{v}=(\frac{\dot{\omega}}{\omega}+\frac{\dot{\omega}\ddot{\omega}\mathrm{std}^2(\boldsymbol{\tau})}{\omega^2})(j\omega\mathbf{x}-j\dot{\omega}\mathbf{Tx})\\&=j\dot{\omega}\mathbf{x}-j\frac{\dot{\omega}^2}{\omega}\mathbf{Tx}+j\frac{\dot{\omega}\ddot{\omega}\mathrm{std}^2(\boldsymbol{\tau})}{\omega}\mathbf{x}-j\frac{\dot{\omega}^2\ddot{\omega}\mathrm{std}^2(\boldsymbol{\tau})}{\omega^2}\mathbf{Tx}\end{aligned} \tag{69}$$

Calculate the normal acceleration $\mathbf{a}_{\perp}=\mathbf{a}_{cc}-\mathbf{a}_{\parallel}$.

$$\begin{aligned}\mathbf{a}_{\perp}&=-\omega^2\mathbf{x}+j\dot{\omega}\mathbf{x}-j\ddot{\omega}\mathbf{Tx}+2\omega\dot{\omega}\mathbf{Tx}-\dot{\omega}^2\mathbf{T}^2\mathbf{x}+...\\&\quad j\frac{\dot{\omega}^2}{\omega}\mathbf{Tx}-j\dot{\omega}\mathbf{x}-j\frac{\dot{\omega}\ddot{\omega}\mathrm{std}^2(\boldsymbol{\tau})}{\omega}\mathbf{x}+j\frac{\dot{\omega}^2\ddot{\omega}\mathrm{std}^2(\boldsymbol{\tau})}{\omega^2}\mathbf{Tx}\\&=\underbrace{(-\omega^2-j\frac{\dot{\omega}\ddot{\omega}\mathrm{std}^2(\boldsymbol{\tau})}{\omega})}_{C_1}\mathbf{x}+\underbrace{(-\dot{\omega}^2)}_{C_3}\mathbf{T}^2\mathbf{x}+...\\&\quad\underbrace{(2\omega\dot{\omega}-j\ddot{\omega}+j\frac{\dot{\omega}^2}{\omega}+j\frac{\dot{\omega}^2\ddot{\omega}\mathrm{std}^2(\boldsymbol{\tau})}{\omega^2})}_{C_2}\mathbf{Tx}\end{aligned} \tag{70}$$

To obtain the magnitude of the normal acceleration, we need to analyze the orthogonality of the vectors $\mathbf{x}$, $\mathbf{Tx}$, and $\mathbf{T}^2\mathbf{x}$. We have

$$\begin{aligned}&\langle\mathbf{x},\mathbf{x}\rangle_{\mathbb{R}}=M\\&\langle\mathbf{Tx},\mathbf{Tx}\rangle_{\mathbb{R}}=M\mathrm{std}^2(\boldsymbol{\tau})\\&\langle\mathbf{T}^2\mathbf{x},\mathbf{T}^2\mathbf{x}\rangle_{\mathbb{R}}=M\mu_4,\mu_4=\frac{1}{M}\sum\boldsymbol{\tau}^4\\&\langle\mathbf{x},\mathbf{Tx}\rangle_{\mathbb{R}}=0\\&\langle\mathbf{Tx},\mathbf{T}^2\mathbf{x}\rangle_{\mathbb{R}}=\sum\boldsymbol{\tau}^3=M\mu_3\approx0\\&\langle\mathbf{x},\mathbf{T}^2\mathbf{x}\rangle_{\mathbb{R}}=\sum1\cdot\boldsymbol{\tau}^2=M\sigma^2\neq0\end{aligned} \tag{71}$$

Therefore, apart from the strong correlation between the reference term $\mathbf{x}$ and the second-order geometric term $\mathbf{T}^2\mathbf{x}$, the magnitude of the other terms can be calculated using the vector sum formula.

$$\|\mathbf{a}_{\perp}\|^2=\|C_1\mathbf{x}\|^2+\|C_2\mathbf{Tx}\|^2+\|C_3\mathbf{T}^2\mathbf{x}\|^2+2\mathrm{Re}\left(C_1^*C_3\langle\mathbf{x},\mathbf{T}^2\mathbf{x}\rangle_{\mathbb{R}}\right) \tag{72}$$

The terms can be calculated as

$$\begin{aligned}&\|C_1\mathbf{x}\|^2=(\omega^4+\frac{\dot{\omega}^2\ddot{\omega}^2\mathrm{std}^4(\boldsymbol{\tau})}{\omega^2})M\\&\|C_2\mathbf{Tx}\|^2=\left(4\omega^2\dot{\omega}^2+(\frac{\dot{\omega}^2}{\omega}-\ddot{\omega}+\frac{\dot{\omega}^2\ddot{\omega}\mathrm{std}^2(\boldsymbol{\tau})}{\omega^2})^2\right)M\mathrm{std}^2(\boldsymbol{\tau})\\&\|C_3\mathbf{T}^2\mathbf{x}\|^2=\dot{\omega}^4M\mu_4\\&2\mathrm{Re}\left(C_1^*C_3\langle\mathbf{x},\mathbf{T}^2\mathbf{x}\rangle_{\mathbb{R}}\right)=2\omega^2\dot{\omega}^2M\mathrm{std}^2(\boldsymbol{\tau})\end{aligned} \tag{73}$$

Under typical radar and communication parameters, terms $M\omega^4$ and $\omega^2\dot{\omega}^2M\mathrm{std}^2(\boldsymbol{\tau})$ contribute the vast majority of the energy. The contributions from the remaining terms differ from the main terms by at least an order of magnitude of $10^{-6}$, thus

$$\|\mathbf{a}_{\perp}\|^2\approx M\omega^4+6\omega^2\dot{\omega}^2M\mathrm{std}^2(\boldsymbol{\tau})+O(\cdots) \tag{74}$$

Assuming $\delta=\frac{\dot{\omega}}{\omega}\mathrm{std}(\tau)$, we can simplify the numerator to

$$\|\mathbf{a}_{\perp}\|\approx\sqrt{M\omega^4(1+6\delta^2)} \tag{75}$$

And the denominator to

$$\|\mathbf{v}\|^2\approx(M\omega^2(1+\delta^2))\approx\sqrt{M^2\omega^4(1+2\delta^2)} \tag{76}$$

Substituting the above results into the curvature formula

$$\kappa_1(t)=\frac{\|\mathbf{a}_{\perp}\|}{\|\mathbf{v}\|^2}\approx\frac{\sqrt{M\omega^4(1+6\delta^2)}}{\sqrt{M^2\omega^4(1+2\delta^2)}}=\sqrt{\frac{1}{M}}\sqrt{1+\frac{4\delta^2}{(1+2\delta^2)}} \tag{77}$$

The complete expression is

$$\kappa_1(\theta,t)\approx\sqrt{\frac{1}{M}}\sqrt{1+\frac{4\dot{\omega}^2\mathrm{std}^2(\tau(\theta))}{\omega^2+2\mathrm{std}^2(\tau(\theta))}} \tag{78}$$

Clearly, there is $\omega^2\gg2\mathrm{std}^2(\tau(\theta))$. Let $\kappa_{\mathrm{geo}}=\frac{1}{\sqrt{M}}$ and $\kappa_{dyn}=\frac{2}{\sqrt{M}}\frac{\dot{\omega}(t)}{\omega(t)}\mathrm{std}(\boldsymbol{\tau}(\theta))$, the first-order curvature can be expressed in a more concise form.

$$\begin{aligned}\kappa_1(\theta,t)&\approx\sqrt{\frac{1}{M}+\left(\frac{2}{\sqrt{M}}\frac{\dot{\omega}(t)}{\omega(t)}\mathrm{std}(\boldsymbol{\tau}(\theta))\right)^2}\\&=\sqrt{\kappa_{\mathrm{geo}}^2+\kappa_{dyn}^2}\end{aligned} \tag{79}$$

**Appendix B: Derivation of the Torsion Projection Formula**

This appendix derives the computational formula for the second-order torsion $\kappa_2$ starting strictly from its differential geometric definition.

To find the torsion, we must calculate the jerk vector $\mathbf{j}(t)=\dot{\mathbf{a}}_{cc}(t)$. We focus on the terms that generate components orthogonal to the osculating plane. A crucial point is that performing higher-order expansions of velocity and acceleration poses computational pitfalls for analytical solutions. As a compromise, we utilize the narrowband assumption to reduce velocity and acceleration to terms involving only the instantaneous frequency itself. Recall the multi-term expansion of the acceleration vector

$$\mathbf{a}_{cc}\approx-\omega^2\mathbf{x}+2\omega\dot{\omega}\mathbf{Tx} \tag{80}$$

Using the derivative rule $\dot{\mathbf{x}}=\mathbf{\Omega x}\approx j\omega\mathbf{x}$, we differentiate the key terms that contribute to the binormal direction.

$$\frac{d}{dt}\mathbf{a}_{cc}\approx-2\omega\dot{\omega}\mathbf{x}-\omega^3j\mathbf{x}+(2\dot{\omega}^2+2\omega\ddot{\omega})T\mathbf{x}+2\omega^2\dot{\omega}j\mathbf{Tx} \tag{81}$$

The osculating plane is spanned by $\mathbf{x}$ (Normal baseline), $j\mathbf{x}$ (Tangent baseline), and $\mathbf{Tx}$ (Normal perturbation). The dominant component of the jerk vector orthogonal to the osculating plane is $j\mathbf{Tx}$.

$$\mathbf{j}_{\perp}\approx2\omega^2\dot{\omega}j\mathbf{Tx} \tag{82}$$

We verify that $\mathbf{j}_{\perp}$ is orthogonal to the osculating plane basis in the real embedded manifold $\mathbb{R}^{2M}$.

The osculating plane is spanned by $\mathbf{v}_{base} \approx j\omega\mathbf{x}$ and Normal baseline $\mathbf{a}_{base} \approx -\omega^2\mathbf{x} + 2\omega\dot{\omega}\mathbf{T}\mathbf{x}$ (Real part). Then we can check the inner product

$$\begin{aligned} \langle \mathbf{x}, j\mathbf{T}\mathbf{x} \rangle_{\mathbb{R}} &= \mathrm{Re}(\mathbf{x}^H j\mathbf{T}\mathbf{x}) = 0 \\ \langle j\mathbf{x}, j\mathbf{T}\mathbf{x} \rangle_{\mathbb{R}} &= \mathrm{Re}(-j\mathbf{x}^H j\mathbf{T}\mathbf{x}) = \mathrm{Re}(\mathbf{x}^H\mathbf{T}\mathbf{x}) = 0 \\ \langle \mathbf{x}T, j\mathbf{T}\mathbf{x} \rangle_{\mathbb{R}} &= \mathrm{Re}(\mathbf{x}^H\mathbf{T}j\mathbf{T}\mathbf{x}) = 0 \end{aligned} \quad (83)$$

Thus, the vector direction $j\mathbf{T}\mathbf{x}$ constitutes the **Binormal Direction** $\mathbf{u}_3$.

Substituting $\mathbf{j}_\perp$ into the torsion formula $\kappa_2 = \dfrac{\|\mathbf{j}_\perp\|}{\kappa_1 \|\mathbf{v}\|^3}$

$$\|\mathbf{j}_\perp\| \approx \left|2\omega^2\dot{\omega}\right| \cdot \|\mathbf{T}\mathbf{x}\| = \left|2\omega^2\dot{\omega}\right| \cdot \sqrt{M}\,\mathrm{std}(\tau) \quad (84)$$

$$\mathrm{Denom} \approx \frac{1}{\sqrt{M}}(M\omega^2)^{3/2} = M\omega^3$$

$$\kappa_2(\theta,t) \approx \frac{|2\omega^2\dot{\omega}(t)|\sqrt{M}\,\mathrm{std}(\tau)}{M\omega^3(t)} \quad (85)$$

Simplifying yields the general torsion law

$$\kappa_2(\theta,t) \approx \frac{2}{\sqrt{M}}\left|\frac{\dot{\omega}(t)}{\omega(t)}\right| \cdot \mathrm{std}(\boldsymbol{\tau}(\theta)) \quad (86)$$